\documentclass{emulateapj} 
 
\newcommand{\lsim}{\raisebox{-0.13cm}{~\shortstack{$<$ \\[-0.07cm] $\sim$}}~}
\newcommand{\gsim}{\raisebox{-0.13cm}{~\shortstack{$>$ \\[-0.07cm] $\sim$}}~}
\newcommand{\tfm}{$\rm 24 \, \mu m \,$}
\newcommand{\nLntfm}{$\nu L_\nu^{\rm 24 \, \mu m} \,$}
\newcommand{\nLnn}{$\nu L_\nu^{\rm 24 \, \mu m} < 10^{10}\, \rm L_{\odot} \,$}
\newcommand{\nLnl}{$10^{10} < \nu L_\nu^{\rm 24 \, \mu m} < 10^{11}\, \rm L_{\odot} \,$}
\newcommand{\nLnu}{$\nu L_\nu^{\rm 24 \, \mu m} > 10^{11}\, \rm  L_{\odot} \,$}
\newcommand{\lb}{$L_{\rm bol.}^{\rm IR}\,$}
\newcommand{\oii}{$\rm [OII] \, \lambda 3727 \,$}
\newcommand{\hb}{$\rm H\beta \, \lambda 4861 \,$}
\newcommand{\oiii}{$\rm [OIII] \, \lambda 5007 \,$}
\newcommand{\ha}{$\rm H\alpha \, \lambda 6563 \,$}
\newcommand{\nii}{$\rm [NII] \, \lambda 6584 \,$}

\shorttitle{The optical spectra of \tfm galaxies in COSMOS: I}
\shortauthors{K. I. Caputi et al.}

\begin{document} 
 
 
\title{The optical spectra of \tfm galaxies in the COSMOS field:  \\
I. {\em Spitzer/MIPS} bright sources in the \lowercase{z}COSMOS-bright 10\lowercase{k} catalogue}

 
\author{K. I. \ Caputi\altaffilmark{1,2}, 
S. J.  Lilly\altaffilmark{1}, 
H.  Aussel\altaffilmark{3},
D. Sanders\altaffilmark{4},
D. Frayer\altaffilmark{5}, 
O. Le F\`evre\altaffilmark{6},
A. Renzini\altaffilmark{7},
G. Zamorani\altaffilmark{8},
M. Scodeggio\altaffilmark{9},
T. Contini\altaffilmark{10},
N. Scoville\altaffilmark{11}, 
C. M. Carollo\altaffilmark{1},
G. Hasinger\altaffilmark{12},
A. Iovino\altaffilmark{13}, 
V. Le Brun\altaffilmark{6},
E. Le Floc'h\altaffilmark{4},
C. Maier\altaffilmark{1}, 
V. Mainieri\altaffilmark{14},
M. Mignoli\altaffilmark{8}, 
M. Salvato\altaffilmark{5},
D. Schiminovich\altaffilmark{15},
J. Silverman\altaffilmark{1,12}, 
J. Surace\altaffilmark{5},
L. Tasca\altaffilmark{6},  
U. Abbas\altaffilmark{6},
S. Bardelli\altaffilmark{8},
M. Bolzonella\altaffilmark{8},
A. Bongiorno\altaffilmark{8},
D. Bottini\altaffilmark{9},
P. Capak\altaffilmark{11}, 
A. Cappi\altaffilmark{8},
P. Cassata\altaffilmark{6},
A. Cimatti\altaffilmark{8},
O. Cucciati\altaffilmark{13},
S. de la Torre\altaffilmark{6},
L. de Ravel\altaffilmark{6},
P. Franzetti\altaffilmark{9},
M. Fumana\altaffilmark{9},
B. Garilli\altaffilmark{9},
C. Halliday\altaffilmark{16},
O. Ilbert\altaffilmark{4},
P. Kampczyk\altaffilmark{1},
J. Kartaltepe\altaffilmark{4},
J.-P. Kneib\altaffilmark{6},
C. Knobel\altaffilmark{1}, 
K. Kovac\altaffilmark{1},
F. Lamareille\altaffilmark{10},
A. Leauthaud\altaffilmark{6},
J. F. Le Borgne\altaffilmark{10},
D. Maccagni\altaffilmark{9},
C. Marinoni\altaffilmark{13},
H. McCracken\altaffilmark{17},
B. Meneux\altaffilmark{9},
P. Oesch \altaffilmark{1},
R. Pell\`o\altaffilmark{10},
E. P\'erez-Montero\altaffilmark{10},
C. Porciani\altaffilmark{1},
E. Ricciardelli\altaffilmark{7},
R. Scaramella\altaffilmark{18},
C. Scarlata\altaffilmark{1},
L. Tresse\altaffilmark{6},
D. Vergani\altaffilmark{9},
J. Walcher\altaffilmark{6},
M. Zamojski\altaffilmark{15},
E. Zucca\altaffilmark{8}
}
\altaffiltext{1}{Institute of Astronomy, Swiss Federal Institute of Technology (ETH H\"onggerberg), CH-8093, Z\"urich, Switzerland.}
\altaffiltext{2}{E-mail address: caputi@phys.ethz.ch}
\altaffiltext{3}{CEA/DSM-CNRS, Universit\'e Paris Diderot, DAPNIA/SAp, Orme des Merisiers, 91191 Gif-sur-Yvette, France.}
\altaffiltext{4}{Institute for Astronomy, University of Hawaii, Honololu, HI, USA.}
\altaffiltext{5}{Spitzer Science Center. California Institute of Technology, Pasadena, CA, USA.}
\altaffiltext{6}{Laboratoire d'Astrophysique de Marseille, France.}
\altaffiltext{7}{Dipartimento di Astronomia, Universit\`a di Padova, Padova, Italy.}
\altaffiltext{8}{INAF Osservatorio Astronomico di Bologna, Bologna, Italy.}
\altaffiltext{9}{INAF-IASF Milano, Milan, Italy.}
\altaffiltext{10}{Laboratoire d'Astrophysique de l'Observatoire Midi-Pyr\'en\'ees, Toulouse, France.}
\altaffiltext{11}{California Institute of Technology, Pasadena, CA, USA.}
\altaffiltext{12}{Max Planck Institut f\"ur Extraterrestrische Physik, Garching, Germany.}
\altaffiltext{13}{INAF Osservatorio Astronomico di Brera, Milano, Italy}
\altaffiltext{14}{European Southern Observatory, Garching, Germany.}
\altaffiltext{15}{Department of Astronomy, Columbia University, MC 2457, 550 West 120th Street, New York, NY 10027, USA.}
\altaffiltext{16}{INAF Osservatorio Astrofisico di Arcetri, Florence, Italy.}
\altaffiltext{17}{Institut d'Astrophysique de Paris, UMR 7095 CNRS, Universit\'e Pierre et Marie Curie, 98 bis Boulevard Arago, F-75014 Paris, France.}
\altaffiltext{18}{INAF, Osservatorio di Roma, Monteporzio Catone (RM), Italy}



\begin{abstract}

  We study zCOSMOS-bright optical spectra for 609 {\em Spitzer/MIPS} \tfm-selected galaxies with $S_{24 \, \rm \mu m}> 0.30 \, \rm mJy$ and $I<22.5$ (AB mag) over 1.5 deg$^2$ of the COSMOS field.  From  emission-line diagnostics we find that: 1)  star-formation rates (SFR) derived from the observed \ha and \hb lines  underestimate, on average, the total SFR by factors $\sim 5$ and 10, respectively; 2) both the Calzetti et al. and the Milky Way reddening laws are suitable to describe the extinction observed in infrared (IR) sources in most cases; 3) some IR galaxies at $z<0.3$  have low abundances, but many others with similar IR luminosities and redshifts are chemically enriched; 4) The average \oiii/\hb ratios of \nLnu galaxies at $0.6<z<0.7$ are $\sim 0.6$ dex higher than the average ratio of all zCOSMOS galaxies at similar redshifts.  Massive star formation and active galactic nuclei (AGN) could simultaneously be present in those galaxies with the highest ionising fluxes;  5)  $\sim 1/3$ of the galaxies with metallicity measurements at $0.5<z<0.7$  lie  below the general mass-metallicity relation at the corresponding redshifts. The strengths of the 4000 $\rm \AA$ break and the $\rm H\delta$ EW of our galaxies show that  secondary bursts of star formation are needed to explain the spectral properties of most IR sources. The LIRG and ULIRG phases occur, on average, between $10^7$ and $10^8$ years after the onset of a starburst on top of underlying older stellar populations. These results are valid for galaxies of different IR luminosities at $0.6<z<1.0$ and seem independent of the mechanisms triggering star formation.
   
\end{abstract}



\section{Introduction}
\label{sec-intro}

Spectroscopic surveys are probably the finest diagnostics to probe the nature of galaxies. They provide accurate redshift determinations and directly reveal the physical and chemical processes governing galaxy evolution through cosmic time. The high cost of spectroscopic campaigns has prevented for many years the follow-up of a significant number of sources in large areas of the sky. Yet, extensive spectroscopic surveys are necessary in order to characterise in detail different galaxy populations in a representative way.

Among galaxy populations, infrared (IR) galaxies have been a subject of major interest in observational cosmology since long before the discovery of the extragalactic IR background (Puget et al.~1996). The most recent determinations indicate that the integrated emission from IR galaxies has a significant contribution to the extragalactic background light, comparable to that produced at optical wavelengths (Dole et al.~2006). As signposts of star-formation and active galactic nucleus (AGN) activity, IR galaxies play a major role in reconstructing galaxy formation, evolution and the history of stellar mass assembly.

Surveys conducted with current-generation IR facilities --such as the {\em Spitzer Space Telescope} (Werner et al.~2004) and now also the {\em Akari Telescope} (Matsuhara et al.~2006)--  are rapidly improving our understanding of the nature and evolution of IR galaxies up to high redshifts. It is now well established that luminous and ultra-luminous IR galaxies (LIRGs and ULIRGs; Sanders \& Mirabel 1996) are much more common at high ($z\gsim0.5$)  redshifts  than in the local Universe and the star formation rate (SFR) density strongly increases up to redshift $z\sim1$ (e.g. Le Floc'h et al.~2005; Caputi et al.~2006a, 2007). These results have been previously suggested by studies conducted with the {\em Infrared Space Observatory (ISO)} (e.g. Aussel et al.~1999; Chary \& Elbaz~2001; Franceschini et al.~2001).   In addition, multiwavelength photometric observations of IR galaxies have produced further information on their different properties:  already-assembled stellar masses (Daddi et al.~2005; Caputi et al.~2006a,b;  Papovich et al.~2006), characteristic spectral energy distributions (SEDs; e.g. Dale et al.~2005;  Rowan-Robinson et al.~2005; Rocca-Volmerange et al.~2007; Takagi et al.~2007; Bavouzet et al.~2008),   and the AGN fraction among IR sources (e.g. Alonso-Herrero et al.~2006; Treister et al.~2006).

Spectroscopic data are necessary to address some other questions, such as e.g. the chemical composition and metallicities of IR galaxies at different redshifts. Also, they are useful to understand the distribution of dust within a galaxy, by comparing the degree of reddening independently inferred from  the IR flux and from the spectral line decrement. Finally, spectroscopic data can help to disentangle the  presence of AGN among IR galaxies.

Some of the first spectroscopic surveys of {\em Infrared Astronomical Satellite (IRAS)} sources studied the nuclear regions of nearby LIRGs and ULIRGs (Kim et al.~1995; Veilleux et al.~1995, 1999). Later, Franceschini et al.~(2003); Flores et al.~(2004) and Liang et al.~(2004) studied the spectra of {\em ISO}-selected IR galaxies at $z\sim0.6-0.7$. More recently, Choi et al.~(2006) analysed spectroscopic star formation rates (SFR) and extinction properties of {\em Spitzer} sources with \tfm detections. Papovich et al.~(2006b) obtained spectroscopic redshifts for a very large sample of \tfm-selected galaxies, but they presented a very limited analysis of the corresponding spectra. Even though all these works investigated different spectral properties of IR sources of different luminosities and redshifts, none of them provides a complete analysis of a large homogeneous sample of IR-selected galaxies from redshifts $z\sim0$ to $z\sim1$. This is the goal of this paper.

 The Cosmic Evolution Survey (COSMOS; Scoville et al.~2007) has been designed to probe galaxy evolution, star formation and the effects of large-scale structure up to high redshifts. The COSMOS field is defined by its {\em Hubble Space Telescope/Advanced Camera for Surveys (HST/ACS)}  2  deg$^2$ coverage. There are multiple follow up observations carried out in the COSMOS field, ranging from X-rays to radio wavelengths. Among the photometric imaging observations, COSMOS includes the full and homogeneous coverage of the field with the {\em Spitzer Infrared Array Camera} ({\em Spitzer/IRAC}; Fazio et al.~2004) and the {\em Multiband Photometer for Spitzer} ({\em MIPS}; Rieke et al.~2004), as part of {\em Spitzer} cycle-2 and 3 Legacy Programs (Sanders et al.~2007).

 COSMOS comprises also a large spectroscopic follow-up program being performed with the {\em Visible Multiobject Spectrograph (VIMOS)} on the {\em Very Large Telescope (VLT)}:  zCOSMOS (Lilly et al.~2007). The zCOSMOS survey consists of two parts: 1) the observation of an $I<$ 22.5 AB mag limited sample of 20,000 galaxies over 1.7 deg$^2$  of the COSMOS field  (zCOSMOS-bright) and 2) a sample of 10,000 galaxies in the central 1 deg$^2$, colour-selected to have redshifts $1.4 < z < 3.0$ (zCOSMOS-deep).

  This paper constitutes the first of a series of studies we are carrying out to analyse the optical spectral properties of \tfm-selected  galaxies in the COSMOS field up to high redshifts. Here, we analyse the sample of  611   \tfm sources with $S_{24 \, \rm \mu m}> 0.30 \, \rm mJy$  
that have been observed so far with zCOSMOS-bright among a total of  $\sim 10,000$ (10k) sources.  This is one of the largest spectroscopic samples of mid-IR-selected galaxies analysed to date in the redshift ranges $0<z\lsim 1$ and $0<z\lsim 3$ for normal galaxies and AGN, respectively. The layout of this paper is as follows: in Section \S\ref{sec_sample}, we describe in detail our datasets and the cross-correlation of \tfm and zCOSMOS-bright sources. In Section \S\ref{sec_rflum}, we compute rest-frame mid-IR luminosities for our galaxies. We present our results on the analysis of the spectra in the following sections: mean and typical dispersion  of the spectra in different redshift and IR luminosity bins (\S\ref{sec_spec}), comparison of different SFR indicators (\S\ref{sec_sfr}), the study of dust properties of our sample (\S\ref{sec_dust}) and line ratio diagnostics and derived metallicities (\S\ref{sec_line}). Later on, in Section \S\ref{sec_agn}, we analyse the AGN present in our sample. In section \S\ref{sec_irph}, we show how the optical spectra can put constraints on the average star formation histories of IR galaxies and, in section \S\ref{sec_young}, we analyse galaxy  candidates for being at the earliest stages of a burst of star formation.    Finally, we summarise and discuss our results in Section \S\ref{sec_concl}.  We adopt throughout a cosmology with $\rm H_0=70 \,{\rm km \, s^{-1} Mpc^{-1}}$, $\rm \Omega_M=0.3$ and $\rm \Omega_\Lambda=0.7$.

\section{Datasets}
\label{sec_sample}
 
\subsection{The zCOSMOS-bright 10k sample} 

 The zCOSMOS survey (Lilly et al.~2007) is a very large spectroscopic program  being performed with VIMOS (Le F\`evre et al.~2003) on the VLT. VIMOS is a  multi-slit spectrograph with four non-contiguous quadrants which cover 7$\times$8 arcmin$^2$ each. zCOSMOS is designed to have a uniform pattern of pointings and a multiple-pass strategy, which guarantee a uniform coverage of the field. We refer the reader to Lilly et al.~(2007) for further technical details about this survey.

  In zCOSMOS-bright, the targets are selected randomly from a complete $I< 22.5$ AB mag catalogue (the `parent catalogue'). This part of the program includes also a set of fainter targets which have been selected for specific reasons, e.g. X-ray or radio sources. All the observations are performed with the $R \sim 600$ VIMOS MR grism, with a spectral coverage of $(5500-9500) \, \rm \AA$.
  
  At the moment of writing, roughly one half of the zCOSMOS-bright targets have already been observed and their spectra have been reduced using the VIPGI software (Scodeggio et al.~2005). Although the data reduction is mostly an automated process, the redshift determination of each source is checked manually on an individual basis by two independent reducers and the results are reconciled later on. The resulting redshift catalogue and set of reduced spectra for this first completed part of zCOSMOS-bright contains 10,644 sources over 1.5 deg$^2$ (`the 10k sample' hereafter). 10,580 out of these sources have $I< 22.5$ AB mag (Lilly et al., in preparation).

  The confidence in the redshift determination of each source is qualified with a flag, whose values can be: 4 (completely secure redshift); 3 (very secure redshift, but with a very marginal possibility of error); 2 (a likely redshift, but with a significant possibility of error); 1 (possible redshift); 0 (no redshift determination); and 9 (redshift based on a single secure narrow line, which is usually \oii or \ha). The addition of $+10$ to the flag indicates a broad-line AGN (BLAGN). 
  
  In addition, photometric redshifts for the entire sample have been obtained with The Zurich Extragalactic Bayesian Redshift Analyzer  (ZEBRA; Feldmann et al.~2006), using the multiple UV through near-IR waveband data available for the COSMOS field.   Tests performed on objects with duplicate spectra of different quality show that good-quality photometric redshifts as those obtained with the COSMOS datasets can be useful to confirm less-secure spectroscopic redshifts, such as those with flag=2, 1 or 9 (see Lilly et al.~2007).

\subsection{The SCOSMOS MIPS shallow \tfm catalogue}

 The SCOSMOS survey (Sanders et al.~2007) comprises {\em IRAC} 3.6, 4.5, 5.8 and 8.0 $\rm \,\mu m$ and {\em MIPS} 24, 70 and 160 $\rm \,\mu m$ observations including the entire 2 deg$^2$ of the COSMOS-ACS field, as part of the {\em Spitzer} cycle-2 and 3 Legacy Programs. In cycle-2, the COSMOS field has been mapped at \tfm down to a completeness flux density $S_{24 \, \rm \mu m}= 0.30 \, \rm mJy$ (SCOSMOS-shallow). A smaller 30$\times$20 arcmin$^2$ region has been observed down to  a flux density limit $S_{24 \, \rm \mu m}\approx 0.08 \, \rm mJy$  as a verification field. The observations performed in cycle-3 extend these deeper maps to the entire  COSMOS area.

\begin{figure}
\epsscale{0.80}
\plotone{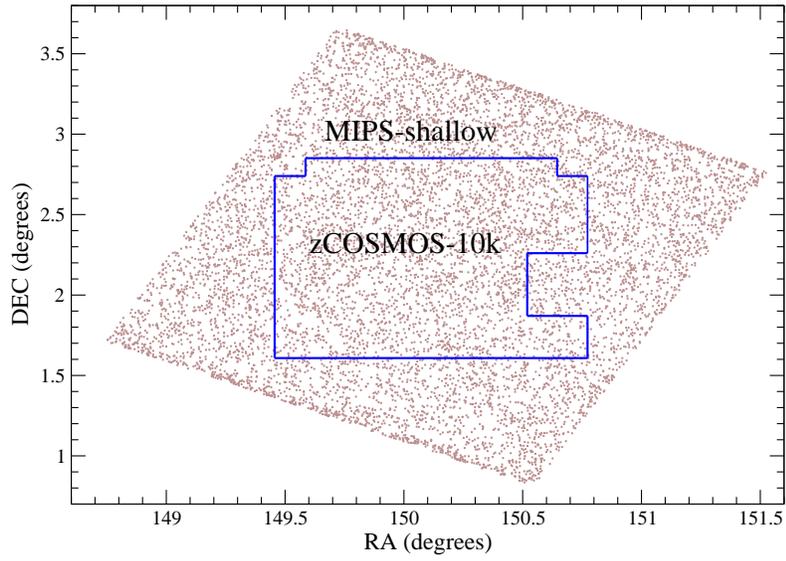}
\caption[]{\label{fig_field} The zCOSMOS-bright 10k and the SCOSMOS/MIPS shallow sample coverage fields.}
\end{figure}

 The source extraction on the \tfm maps has been performed with the IDL version of the DAOPHOT package. A point spread function (PSF) fitting technique for measuring the photometry has been  necessary to deal with blending problems. Further details on the SCOSMOS survey as well as the data reduction and source detection procedures are given by Sanders et al.~(2007). The final SCOSMOS-shallow \tfm catalogue contains 9,807  sources with $S_{24 \, \rm \mu m}> 0.30 \, \rm mJy$  over an area 1.75$\times$1.97 deg$^2$. Out of them, 3,150 sources lie within the zCOSMOS-bright 10k-sample field (see Figure \ref{fig_field}).
 
We explain in detail the selection effect produced by the zCOSMOS-bright 10k sample in next section and summarise the numbers of sources in Table \ref{tab_galsel}.

\subsection{Cross-correlation of the catalogues}

We firstly checked  the fraction of \tfm sources present in the zCOSMOS-bright parent catalogue, i.e. the catalogue from which the spectroscopic targets are taken randomly. This is necessary to assess how representative the zCOSMOS-bright 10k sample is for the identification of  sources in SCOSMOS-shallow. 

\begin{figure}
\epsscale{0.80}
\plotone{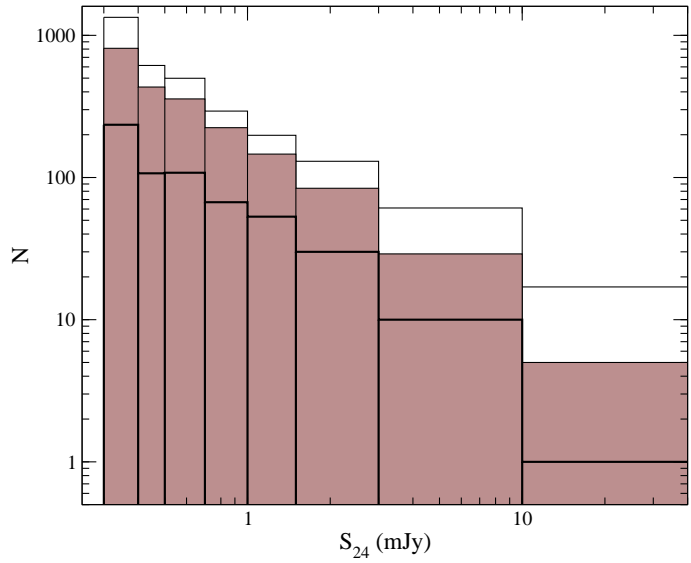}
\caption[]{\label{fig_flhisto} Distributions of \tfm fluxes of all the SCOSMOS-shallow \tfm sources present in the zCOSMOS-bright 10k-sample field (empty histogram); those that have a counterpart in the zCOSMOS-bright parent catalogue (thin-line, shaded histogram); and those that have a good-quality redshift in the zCOSMOS-bright 10k catalogue (thick-line, shaded histogram).
}
\end{figure}

We cross-correlated the SCOSMOS-shallow \tfm catalogue with the zCOSMOS-bright parent catalogue, using a matching radius of 2 arcsec. We found that 2,084 out of 3,150 ($\sim$66\%) \tfm sources have a counterpart in the zCOSMOS-bright parent catalogue limited to $I<22.5$ AB mag. Four additional sources have fainter $I$-band counterparts which are also in the potential list of targets of zCOSMOS-bright. The \tfm sources with associations in the zCOSMOS-bright parent catalogue have a similar \tfm flux distribution as the entire sample of 3,150  \tfm sources lying in the  zCOSMOS-bright 10k-sample field  (see Figure \ref{fig_flhisto}). Thus, the \tfm sources identified with zCOSMOS-bright objects can be considered as representative of the majority ($\sim 66\%$) of the $S_{24 \, \rm \mu m}> 0.30 \, \rm mJy$ population at different fluxes.

The \tfm sources without an association in the zCOSMOS-bright parent catalogue have $I>22.5$ AB mag. Photometric redshifts obtained with ZEBRA for $I<24$ AB mag sources (Oesch et al., in preparation) indicate that the majority ($\gsim 80\%$) of the MIPS-shallow sources not present in the zCOSMOS-bright parent catalogue are at $z>1$.

703 out of the  3,150  \tfm sources in the zCOSMOS-bright field have been observed so far and form part of the  zCOSMOS-bright 10k sample.   

Multiple associations are a relatively minor problem in the identification of \tfm sources. To verify the reliability of each zCOSMOS-bright counterpart, we cross-correlated the \tfm catalogue with a general $I<25$ AB mag catalogue available for the COSMOS field (Capak et al.~2007).   In 35 out of 703 cases ($<5\%$), we found that there is an  $I<25$ mag source which is closer to the \tfm centroid than the zCOSMOS-bright associated source. We assumed that the \tfm-zCOSMOS association was false in these cases, and we have not further considered these sources in our analysis.  In other 73  out of 703 cases ($\sim 10\%$), there is an $I<25$ mag source which is also within 2 arcsec distance of the \tfm centroid, but is further away than the zCOSMOS counterpart. We considered these \tfm-zCOSMOS associations as `likely'. The inclusion of these sources does not introduce any significant change in our results. Finally, 595 out of 703 ($\sim 85\%$) are secure one-to-one  \tfm-zCOSMOS associations.  The resulting sampling rate  for the \tfm galaxies with $I<22.5$ AB mag is $\sim 1/3$, similar to the overall sampling rate of the parent catalogue achieved so far with the total 10k sample.

Having 668 \tfm sources with secure or likely zCOSMOS-bright-10k associations, we had to determine how many of them had good-quality zCOSMOS spectroscopic redshifts and spectra. We found that 482 out of these 668 sources have  a zCOSMOS redshift with quality flags 3 or 4.  Additional 129 sources have  less secure spectroscopic redshifts (flags=2,1 or 9), but which are confirmed by ZEBRA photometric redshifts. The analysis of a subset of repeated zCOSMOS spectra of different quality has showed that $> 90\%$ of the spectroscopic redshifts with flag=2,1 or 9 which are confirmed by ZEBRA are correct (Lilly et al., in preparation).

In total,  611 out of the 668 (i.e.  $>90\%$) considered associations have secure redshifts and reasonable-quality  spectra to perform line measurements. Their \tfm-flux distribution is similar to that of all galaxies identified in the zCOSMOS-bright parent catalogue (compare thin and thick-line shaded histograms in Figure \ref{fig_flhisto}). 2 out of the 611 sources with secure redshifts have been identified with galactic stars. The remaining 609 galaxies constitute the sample analysed in this work.
 
The photometric redshifts obtained with ZEBRA for the 57 spectroscopic failures suggest that 70 and 30\% of them are at $z<1$ and $z>1$, respectively. 
  
\section{Rest-frame \tfm luminosities}
\label{sec_rflum}

\begin{figure}
\plotone{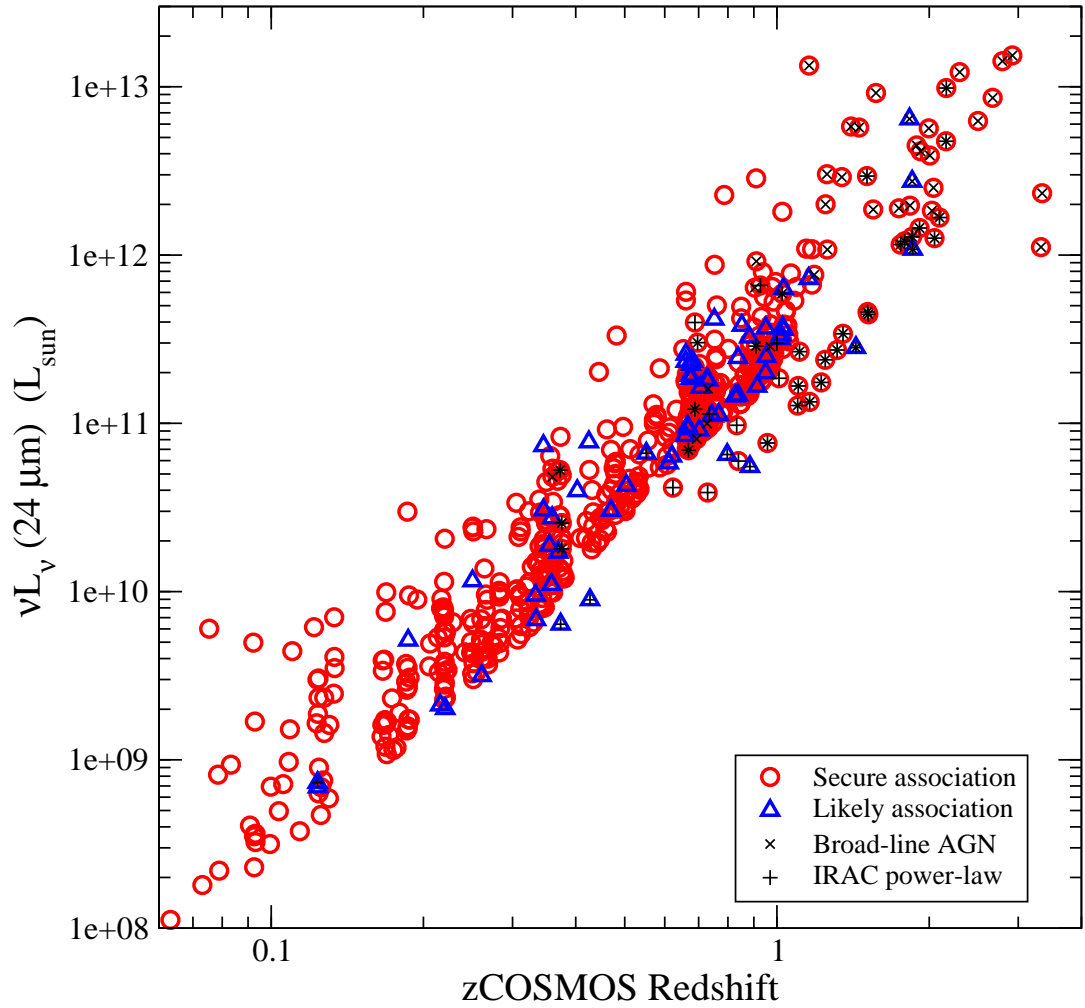}
\caption[]{\label{fig_l24} Rest-frame \tfm luminosities of the 609 galaxies in our zCOSMOS-10k sample versus the spectroscopic redshifts. Red circles correspond to secure one-to-one associations and blue triangles indicate the cases in which there is more than one optical counterpart to the \tfm source within 2 arcsec radius (with the zCOSMOS source being the closest to the \tfm centroid). Crosses and plus-like symbols indicate spectroscopically classified BLAGN and {\em IRAC} power-law SED galaxies, respectively.}
\end{figure}

 We classify the  609 \tfm galaxies in our spectroscopic sample according to their rest-frame IR luminosities and zCOSMOS redshifts. We computed rest-frame \tfm luminosities \nLntfm using the observed \tfm fluxes, the zCOSMOS redshifts and model-derived k-corrections; i.e. \nLntfm$=\nu 4 \pi k(\lambda) S_{24 \, \rm \mu m} d_{\rm L}^2(z)$, where $k(\lambda)$ is the k-correction, $S_{24 \, \rm \mu m}$ is the \tfm flux and $d_{\rm L}(z)$ is the luminosity distance.    We show  the rest-frame \tfm luminosities versus zCOSMOS redshifts for our galaxy sample in Figure \ref{fig_l24}.
 
We used the Lagache et al.~(2003, 2004) IR SED  models to compute the mid-IR k-corrections. As most of our galaxies lie at redshifts $z<1$, the use of other models (e.g. Chary \& Elbaz 2001; Dale \& Helou 2002) would produce rest-frame \tfm luminosities which are consistent within a factor 2. The discrepancies can be considerably larger at higher redshifts (Caputi et al.~2007).

Traditionally, it is preferred to classify IR galaxies according to their bolometric IR (5-1000 $\rm \mu m$) luminosity \lb. However, we decided to classify our sample using their rest-frame \tfm luminosities  \nLntfm, as the monochromatic\footnote{In our case, the rest-frame IR luminosities are convolved with the \tfm filter response function rather than monochromatic.} IR luminosity is a less model-dependent quantity.  As we explain below, we do apply recipes to convert the rest-frame \nLntfm to bolometric IR luminosities when necessary, e.g. for the computation of star formation rates (SFR). Roughly, a factor $\sim10$ should be applied to convert \nLntfm into \lb (although this factor decreases with increasing \nLntfm). Thus, galaxies with  \nLnn, \nLnl and \nLnu  would roughly correspond to IR normal galaxies, LIRGs and ULIRGs, respectively.

The Lagache et al. and other SED models are applicable to IR galaxies dominated by star formation. The SED of some  AGN, instead, have a power-law shape $f_{\nu} \propto \nu^\alpha$  ($\alpha<0$; see e.g. Elvis et al.~1994; Alonso-Herrero et al.~2006). 

We investigated the presence of IR power-law SED galaxies within our sample, by analysing the corresponding {\em Spitzer/IRAC} photometry. We classified as {\em IRAC} power-law sources those galaxies having fluxes in the four {\em IRAC} channels 3.6, 4.5, 5.8 and 8.0 $\rm \mu m$ that are consistent  with a single power law $f_{\nu} \propto \nu^\alpha \, (\alpha<0)$, within the error bars. For this, we considered a minimum error bar of 0.10 mag in each {\em IRAC} channel.

45 out of the 609 \tfm galaxies in our sample have  {\em IRAC} power-law SED (see Figure \ref{fig_l24}). 39 out of the 45 {\em IRAC} power-law SED sources are detected in the {\em XMM-Newton} X-rays maps of the COSMOS field (Hasinger et al.~2007) and 30 are BLAGN, as classified from their zCOSMOS spectra. We further discuss the {\em IRAC} power-law sources and other  AGN present in our sample in Section \S\ref{sec_agn}.

\section{The spectra}
\label{sec_spec}

Before performing different spectral line diagnostics, we did some general  inspection of the  qualitative and quantitative spectral characteristics of our \tfm galaxies.

Each individual zCOSMOS spectrum covers the observed wavelength range $(5500-9500) \, \rm \AA$ and has a dispersion of $2.55 \, \rm \AA$. The quality of the spectra is quite uniform  up to wavelengths $\lambda \sim 8000 \rm \AA$. At longer wavelengths, the quality is degraded by fringing, which can become quite severe in some cases beyond   $\lambda \sim 8500 \rm \AA$. As we explain later, we only perform line measurements on individual spectra at $\lambda < 8500 \rm \AA$, and all the measurements between  $\lambda=8000$ and $\lambda=8500 \, \rm \AA$ have been manually checked on an individual basis.

\begin{figure}
\epsscale{0.50}
\plotone{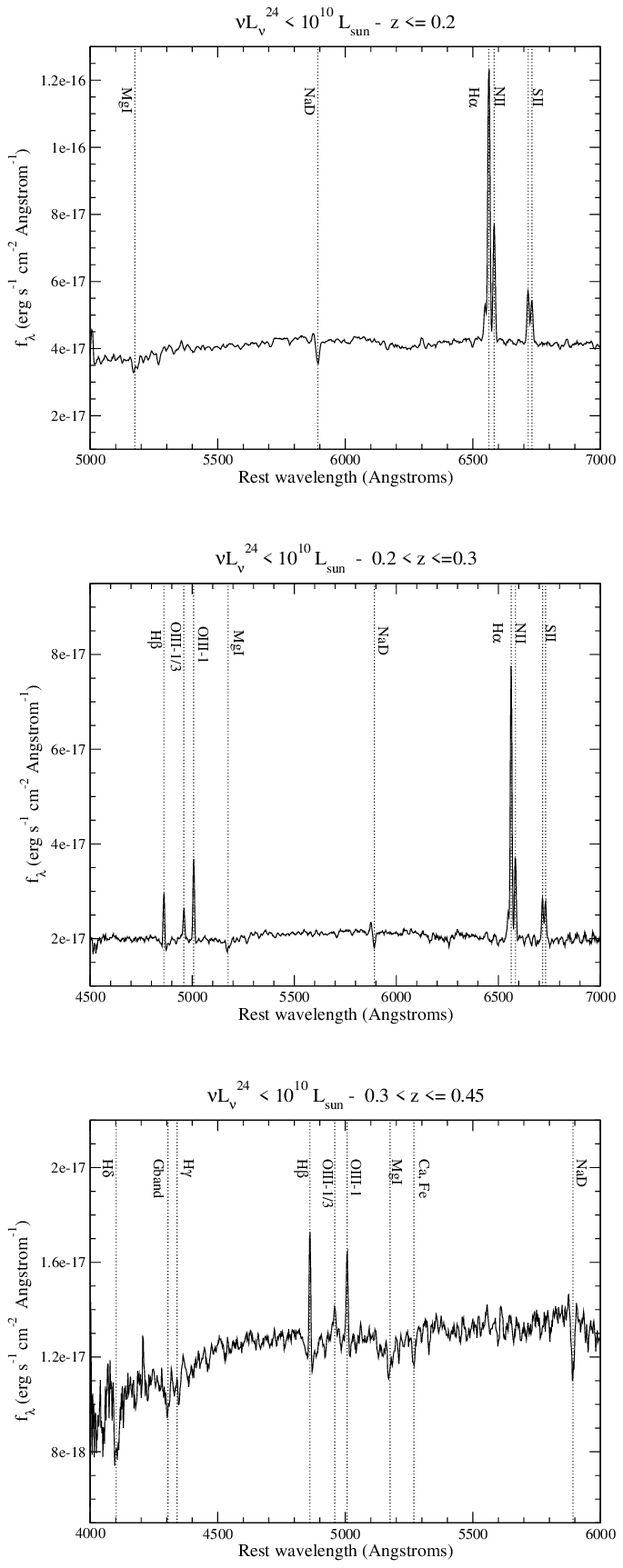}
\caption[]{\label{fig_stack1} Composite average zCOSMOS spectra of \tfm galaxies with \nLnn in different redshift bins. From top to bottom, the numbers of stacked galaxies are 74, 64 and 38, respectively.}
\end{figure}

\begin{figure}
\epsscale{0.50}
\plotone{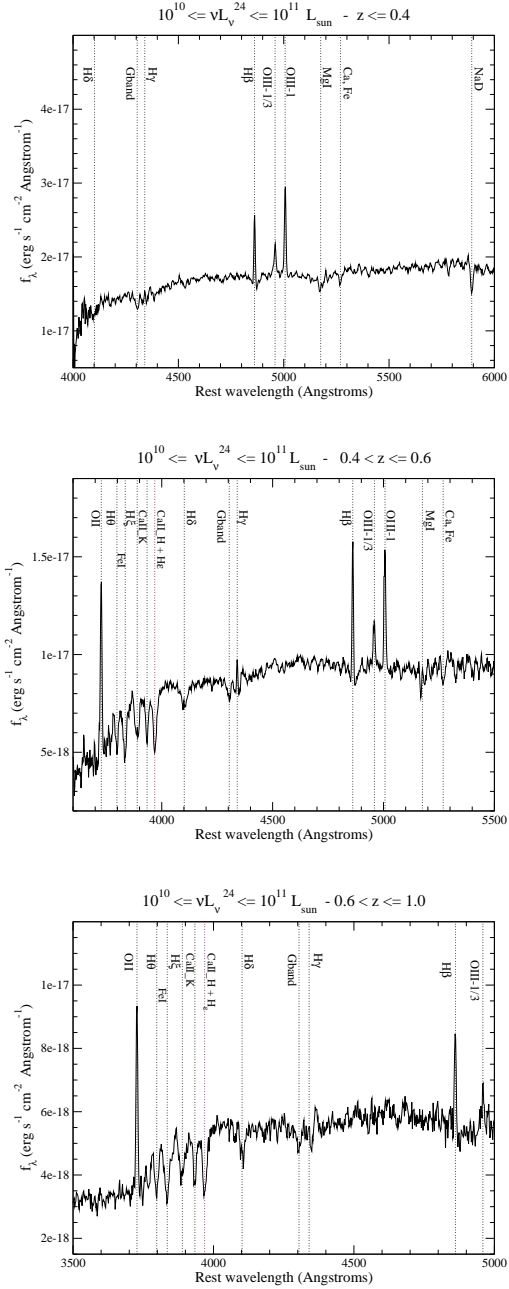}
\caption[]{\label{fig_stack2} Composite average zCOSMOS spectra of \tfm galaxies with \nLnl in different redshift bins. From top to bottom, the numbers of stacked galaxies are 70, 57 and 42, respectively.}
\end{figure}

\begin{figure}
\epsscale{0.50}
\plotone{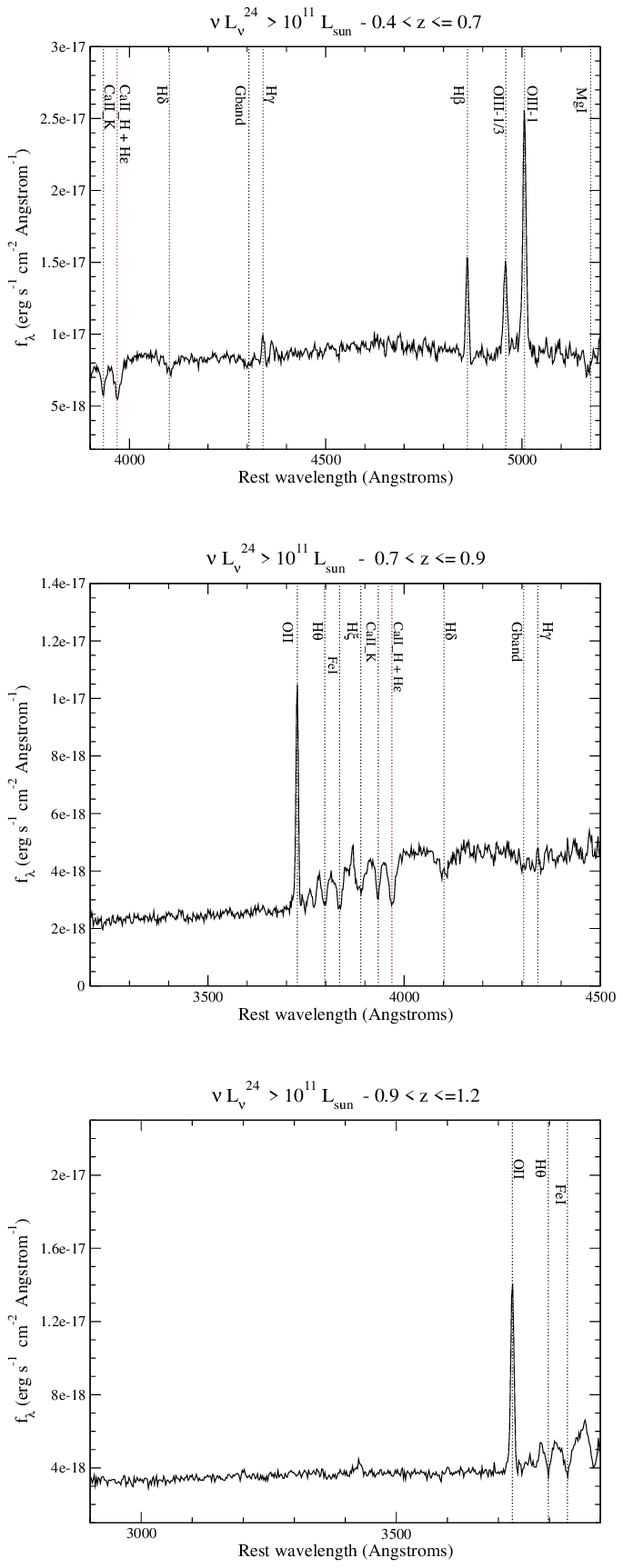}
\caption[]{\label{fig_stack3} Composite average zCOSMOS spectra of \tfm galaxies with \nLnu in different redshift bins. From top to bottom, the numbers of stacked galaxies are 42, 86 and 89, respectively.}
\end{figure}

Stacked spectra can give a good qualitative idea of the mean optical spectra of \tfm galaxies. 
Figure \ref{fig_stack1}, \ref{fig_stack2} and \ref{fig_stack3} show the rest-frame average stacked spectra of our \tfm galaxies with  \nLnn, \nLnl and \nLnu, respectively,  in different redshift bins. We constructed the stacked spectra on sets of 38 to 89 galaxies, depending on the redshift and IR luminosity bin. The zCOSMOS BLAGN have been excluded for the stacking. The wavelength resolution in all the stacked spectra is $\rm 2\AA$ (rest-frame). We only show the spectra up to rest-frame wavelengths corresponding to observed $\rm \lambda \lsim 8500 \, \AA$, i.e. we excluded the regions most affected by fringing. However, as we show in Section \S\ref{sec_line}, even the longest wavelength regions show reasonably good average spectra when sufficient numbers of sources are stacked. This is due to the fact that the fringing pattern adds incoherently for different sources, so no systematic noise is propagated into the average stacked spectra.

We obtained the average spectrum in each  bin  by re-normalising the individual rest-frame spectrum of each source to the average value of a featureless region of the continuum  (which was chosen depending on the redshift bin). In this way, all the individual spectra in a given redshift and IR luminosity bin are put on a same scale before stacking. We smoothed out regions lying on top of the main atmospheric absorption lines, except when source emission lines were present.  At each rest-frame wavelength, we excluded  the 5\% smallest and largest values before computing the average. This sigma-clipping procedure helps to clean the stacks for possible remaining spurious lines in the individual spectra.

These average spectra of \tfm galaxies in different redshift and IR luminosity bins show that:

\begin{itemize}

\item as expected, all emission lines characterising star-forming galaxies are present. We note that these emission lines are a property of the average spectra of IR galaxies, but we do not necessarily  observe all these lines in every spectrum on an individual basis (because e.g. they are much extincted by reddening in some cases). On the average spectra, we also see some absorption lines characteristic of old stellar populations, as NaD or CaII H \& K. This means that different generations of stars are present in many IR galaxies.

\item the average line ratios vary as a function of IR luminosity and redshift. We further analyse this point in Section \S\ref{sec_line}.

\item high-order Balmer absorption lines are clearly present in \tfm galaxies. These lines are produced by short-living ($\rm \lsim 1 \, Gyr$)  A-type stars. This indicates that \tfm galaxies not only are instantaneously forming stars, but have been also forming stars for some time during the last Gyr. We explore this issue in more detail in Section \S\ref{sec_irph}.

\end{itemize}

Although the stacked spectra give a good qualitative idea of the average spectral properties of \tfm galaxies of different IR luminosities and redshifts, they do not contain information about the variety of strengths of the spectral features among  \tfm  galaxies in a same redshift and luminosity bin.

We measured the fluxes and equivalent widths (EW) of emission lines in our zCOSMOS spectra by direct integration on  the rest-frame spectra. We do not measure emission lines lying beyond observer-frame $\lambda=8500 \, \rm \AA$, to avoid being severely affected by fringing. For lines at observer-frame $8000 < \lambda < 8500 \, \rm \AA$, we manually checked the measurements on an individual basis using the IRAF package `splot'. We also checked manually those lines lying on top of or near strong atmospheric absorption features, e.g. \oii at $z \sim 0.7$. In all the following, except when otherwise stated, we only present and analyse line measurements when the corresponding rest-frame EW are  $> 5 \, \rm \AA$,  as those are the lines with the most secure measurements.

The error bars have been computed as the sum of two components. The first error-component has been obtained by considering the variations of the continuum in a narrow region around the line and setting this to a minimum  of 5\%.  On the other hand,  $\sim 10\%$ of the sources on which we performed line measurements have duplicate good-quality spectra. We used these repeated line measurements to work out a second component for the error bars. The error bars computed in the first step have been then systematically shifted, in such a way that their median values coincided with the medians derived from the analysis of duplicates. The systematic percentual errors we added to the line fluxes are 12, 15, 12, 9 and 12\% for \oii, \hb, \oiii, \ha and \nii, respectively. For the EW measurements, the systematic percentual errors added are 17, 20, 17, 10 and 18\%.

All Balmer emission lines have been corrected for stellar absorption. To do this, we fitted the continuum of each spectrum with synthetic stellar SED from the Bruzual \& Charlot 2007 library (Bruzual \& Charlot 2003; Bruzual 2007). The medians of the stellar absorption corrections for \ha and \hb in our galaxies are, respectively, $(1.8\pm0.5) \, \rm \AA$ and $(3.1\pm0.3) \, \rm \AA$.

In addition, we applied aperture corrections to all the spectral lines, as the slits in the VIMOS masks have a width of 1 arcsec. We convolved each zCOSMOS spectrum in our sample with the SUBARU $R$ and $I$-band filters and compared the resulting magnitudes with the total magnitudes of our galaxies, as available in the COSMOS optical photometry catalogues (Capak et al.~2007).  The difference between the two (spectrum-derived and total) magnitudes gave us the aperture correction in each case. For the majority ($>75\%$) of our galaxies, the optical-flux correction factors are between 1 and 4. For less than 2\% of the sample, in all cases $z<0.3$ galaxies, the correction factors can be as large as a factor 10 to 20.

\begin{figure}
\epsscale{0.9}
\plotone{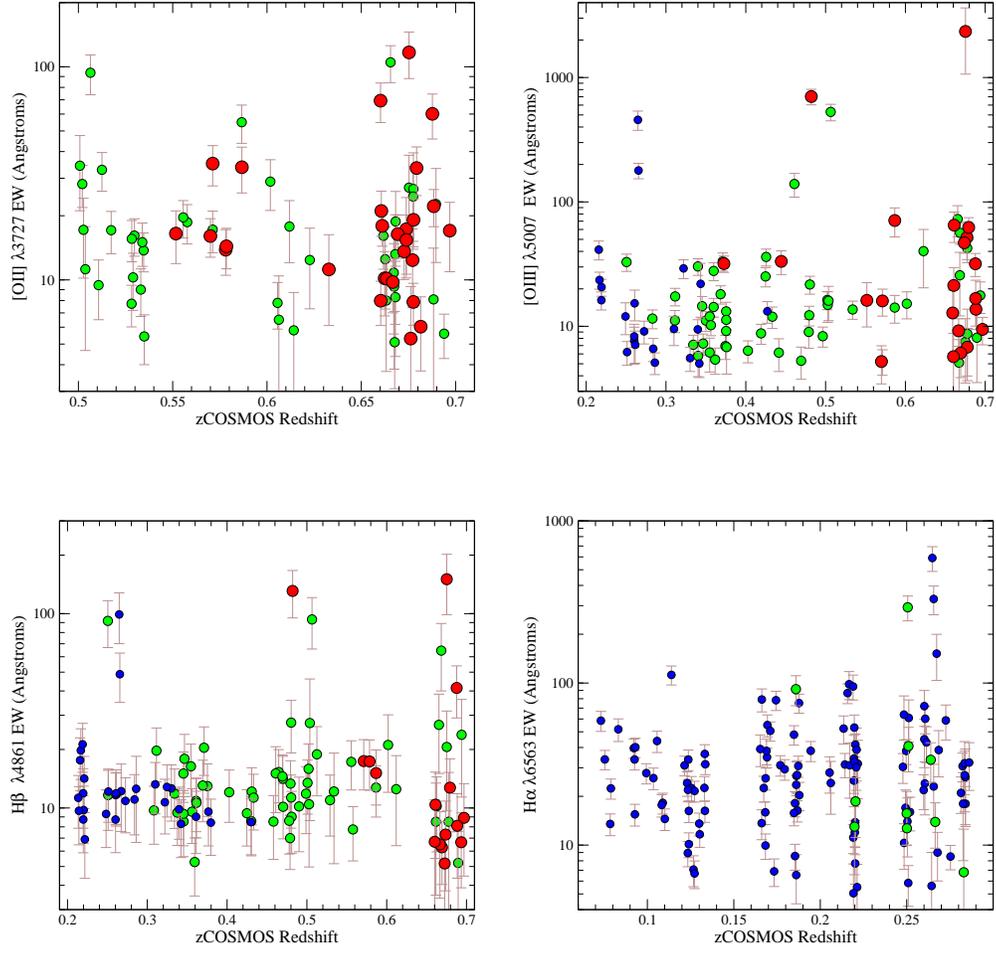}
\caption[]{\label{fig_ewc} Rest-frame EW measurements of different lines in the zCOSMOS spectra of \tfm galaxies as a function of redshift. Symbols of different size and colour correspond to galaxies with different mid-IR luminosities:  \nLnn (small blue circles), \nLnl (medium-size green circles) and \nLnu (large red circles). }
\end{figure}

Figure \ref{fig_ewc} shows the EW of typical emission lines in the spectra of \tfm galaxies as a function of redshift. Only cases with EW $\rm >5 \,\AA$ are presented in this figure. Sources with \ha EW $\rm >5 \,\AA$ are more than 80\% of all our sources at $z<0.3$ (the redshift bin in which we performed \ha measurements on an individual basis). Instead, sources with \hb and \oiii EW $\rm >5 \,\AA$ constitute $\sim$ 35 to 40\% of all our sources at $0.2<z<0.7$. 

From figure \ref{fig_ewc}, we see that the spectral lines of \tfm galaxies display a large range of EW values. Even galaxies with comparable IR luminosities at similar redshifts have  lines whose EW differ, in some cases, by a factor 5 or more. 

There is a wide range of possible \ha EW values for galaxies with \nLnn up to $z=0.3$. This is probably a spatial sampling effect, because the VIMOS slit only samples a small region of 1 arcsec width. Within a star-forming galaxy, different regions such as HII nuclei, hotspots or disk HII regions are characterised by different \ha EW distributions (e.g. Kennicutt et al.~1989). At low redshifts, these different regions can be resolved and they can individually dominate the optical light collected in the 1 arcsec-wide slit. In contrast, given the size of the {\em MIPS} PSF at \tfm ($\sim 6$ arcsec), the IR luminosity corresponds to the  IR light in the integrated galaxy.  The large dispersion in EW is not seen anymore for \hb at higher redshifts, where the different regions cannot be resolved (see Figure \ref{fig_ewc}). However, the \oiii EW do span a wide range of values at these higher redshifts, indicating the existence and different importance of ionising sources in IR galaxies.

\section{Comparison of star formation rate indicators}
\label{sec_sfr}

When all the ultra-violet (UV) light produced by young stars in a  star-forming galaxy is absorbed and re-emitted by the surrounding dust, the IR bolometric ($5-1000 \, \rm \mu m$) luminosity \lb is directly related to the SFR (Kennicutt 1998).   Sometimes, however, part of the light produced by young stars is not absorbed and can directly be observed at UV wavelengths (for a discussion see e.g., Buat et al.~2007). This mainly depends on the geometry and mixing of the dust with young stars. Thus, an optimal estimator of SFR should take into account both the IR and UV contributions.

Our aim is to compare the SFR derived from emission lines with no dust-correction in the zCOSMOS spectra, with the fiducial total (IR+UV) SFR. This allows us, in the next section,  to quantify the degree of extinction characterising \tfm galaxies.

We consider here that the total SFR of our \tfm galaxies is given by the sum of the IR and UV-derived SFR:

\begin{eqnarray}
SFR_{\rm IR} \, (\rm M_\odot \, yr^{-1})&=& 1.72 \times 10^{-10} \times L_{\rm bol.}^{\rm IR}, \\
SFR_{\rm UV} \, (\rm M_\odot \, yr^{-1})&=& 1.4 \times 10^{-28} \times L_\nu^{\rm UV};
\end{eqnarray}

\noindent where \lb is the bolometric IR luminosity expressed in $\rm L_\odot$ and $L_\nu^{\rm UV}$ is the UV luminosity expressed in erg \, s$^{-1}$ Hz$^{-1}$, which is approximately constant  between 1500 and 2800 $\rm \AA$  for a Salpeter (1955) initial mass function (IMF; see Kennicutt 1998). Both equations assume a Salpeter IMF over stellar masses (0.1-100) M$_\odot$.

We computed the bolometric IR luminosity of our galaxies based on the rest-frame \tfm luminosities \nLntfm and using the Bavouzet et al.~(2008) \nLntfm-\lb relation:

\begin{equation}
L_{\rm bol.}^{\rm IR}=6856 \times (\nu L_\nu^{\rm 24 \, \mu m})^{0.71}_{\rm rest} \,\,\, (\pm 54\%).
\end{equation}

\noindent This relation has been calibrated on a sample of {\em Spitzer/MIPS} galaxies detected at 24, 70 and 160 $\rm \mu m$ with spectroscopic redshifts $z<0.6$. The k-corrections of the rest-frame \tfm luminosities used to calibrate this formula depend on galaxy IR SED models. However, in contrast to other typical recipes to do the \nLntfm-\lb conversion, the bolometric IR luminosities used in the Bavouzet et al.~(2008) calibration have been directly calculated from the observed fluxes at 24, 70 and 160  $\rm \mu m$.

We obtained the UV luminosity at a reference rest-frame wavelength  $\lambda=2000 \, \rm \AA$ for our galaxies using the GALEX near-UV (NUV) data for the COSMOS field (Schiminovich et al.~2007), which correspond to an observed effective wavelength $\lambda_{\rm eff.}=2310 \, \rm \AA$. We used the Bruzual \& Charlot (Bruzual~2007) models to compute the corresponding k-corrections. We note that, for most of our galaxies,  the derived UV SFR are a minor fraction of the total SFR ($<10\%$ for $\sim$85\% of our star-forming galaxies and $<20\%$ for $\sim$95\% of them).

 Several emission lines are commonly used in the literature to trace star formation activity, even though all of them are known to present different caveats. Balmer lines are affected by underlying stellar  absorption and dust extinction. \oii and  \oiii are sensitive not only to dust extinction, but also to chemical abundance and the degree of ionisation in the star-forming galaxy (see e.g. Moustakas et al.~2006). To avoid dealing with the additional uncertainties associated with the SFR derived from ionisation lines, we only study here the \ha and \hb-based SFR,  for a comparison with the total (IR+UV) SFR obtained for the \tfm galaxies.

We computed the SFR derived from the \ha line for galaxies with zCOSMOS redshifts $z<0.3$ with the  Kennicutt (1998) formula:

\begin{equation}
\label{eq_ha}
SFR_{\rm H\alpha} \, (\rm M_\odot \, yr^{-1})= 7.94 \times 10^{-42} \times L_{\rm H\alpha},
\end{equation}

\noindent where $L_{\rm H\alpha}$ is the luminosity associated with the flux of the \ha line expressed in erg s$^{-1}$. This formula is also valid for a Salpeter IMF over stellar masses (0.1-100) M$_\odot$.  For galaxies with redshifts $0.2<z<0.7$, we computed the  SFR derived from the \hb line simply using equation (\ref{eq_ha}) and assuming a case B recombination with temperature $T=10,000 \rm K$, i.e an intrinsic ratio (\ha/\hb)$_{\rm int.}$=2.87 (Osterbrock 1989).

\begin{figure}
\epsscale{0.9}
\plotone{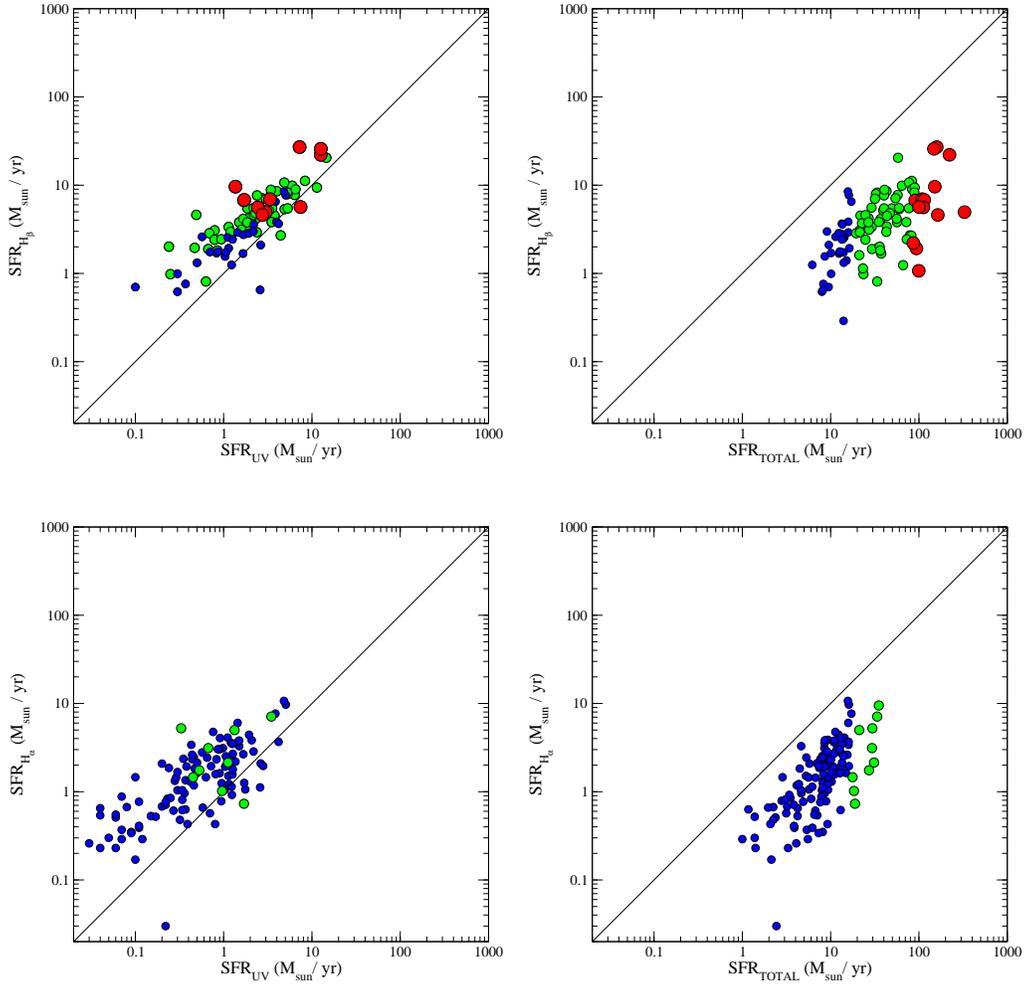}
\caption[]{\label{fig_sfr} Comparison of different SFR indicators: \hb versus UV (top-left) and TOTAL=IR+UV (top-right panel); \ha versus UV (bottom-left) and TOTAL=IR+UV (bottom-right panel). The symbol code is the same as in figure \ref{fig_ewc}. }
\end{figure}

The four panels in figure \ref{fig_sfr} show the SFR derived from \ha and \hb line luminosities compared to the UV only and total (IR+UV) SFR. Galaxies identified with BLAGN from the optical spectra have been excluded from this analysis. We also have excluded 5 other  sources with \ha and/or \hb measurements which are obvious cases of AGN, as determined from their high X-ray luminosities $L_{X}>10^{42.5} \, \rm erg \, s^{-1}$ (see Section \ref{sec_agn}). We performed the comparison with the UV and (IR+UV) SFR  before correcting  the spectral lines for dust, and only including stellar absorption and aperture corrections. This allows us  to understand the differences in the derived SFR because of dust extinction and to assess how reliable the SFR obtained from the extincted \ha and \hb lines are.   In Section \ref{sec_dust}, we derive dust extinction corrections for our galaxies by imposing that the \ha or \hb SFR are equal to the total (IR+UV) SFR. The dust extinction corrections have then been applied for all the subsequent analysis made in the following sections. 

Several conclusions can be extracted from inspection of figure \ref{fig_sfr}. Firstly, the general trend observed is that the derived SFR are $SFR_{\rm UV} < SFR_{\rm line} < SFR_{\rm (IR+UV)}$. As reddening is larger at UV than at optical wavelengths (and even larger than at IR wavelengths), this trend appears to be a consequence of the differential dust extinction.
This is supported by the fact that the SFR obtained from \hb are in better agreement with the UV-only SFR than those derived from \ha (see  left-hand side panels of figure \ref{fig_sfr}).

For galaxies with \nLnn at $z<0.3$, the observed \ha line produces a SFR that is typically a factor $\sim$5 below the total (IR+UV) SFR. The differences are more dramatic when using the \hb line: for galaxies with \nLnl at $0.2<z<0.7$, the SFR are underestimated by a factor $\sim 10$. This indicates that the effects of dust extinction are very important in most IR sources. This is consistent with the results of previous spectroscopy studies of IR-luminous galaxies (e.g.  Flores et al.~2004; Liang et al.~2004; Choi et al.~2006). The exact factors obtained for the optical-to-total SFR conversion mainly depend on the recipes adopted to convert mid-IR into bolometric IR luminosities.

For some galaxies in our sample, especially among the \nLnu population,  the discrepancies between the \hb-derived and total SFR are much larger than the average ($\gsim 50$). This could indicate the presence of either very-obscured star-forming galaxies or the presence of AGN which are to some extent responsible for the IR emission.  We note that, although we excluded from this analysis all the zCOSMOS identified BLAGN, narrow-line AGN or mixed AGN/star-forming systems might still be present in our sample. We mentioned in Section \ref{sec_rflum} the existence of some  {\em IRAC} power-law sources within our sample. However, only one of the sources with highly discrepant SFR is actually an {\em IRAC} power-law. In some other cases,  high ionisation line ratios are observed in their optical spectra. As we will discuss in Section \ref{sec_young}, star formation and AGN activity might coexist in some of these systems.

\section{Dust extinction}
\label{sec_dust}

The discrepancies between the SFR derived from the emission lines in the optical spectra and the (IR+UV) indicators  make necessary the study of dust extinction in our galaxies. The common procedure in the literature is to adopt a known reddening law to deredden the optical spectra. Either the Calzetti et al.~(2000), the Milky Way (MW; Fitzpatrick 1999 and references therein) or the Small Magellanic Cloud (SMC; Pr\'evot et al.~1984) reddening laws are generally used. We can test the adequacy of these different laws to describe the interstellar extinction in a sub-sample of our galaxies.

We analysed the IR, UV and zCOSMOS spectroscopic data of those galaxies in our sample for which both \ha and \hb were present and had sufficiently reliable measurements, i.e. we restricted the analysis  to the 24 \tfm galaxies at $0.2<z<0.3$ with EW$>5 \, \rm \AA$ for both \ha and \hb.

\begin{figure}
\plotone{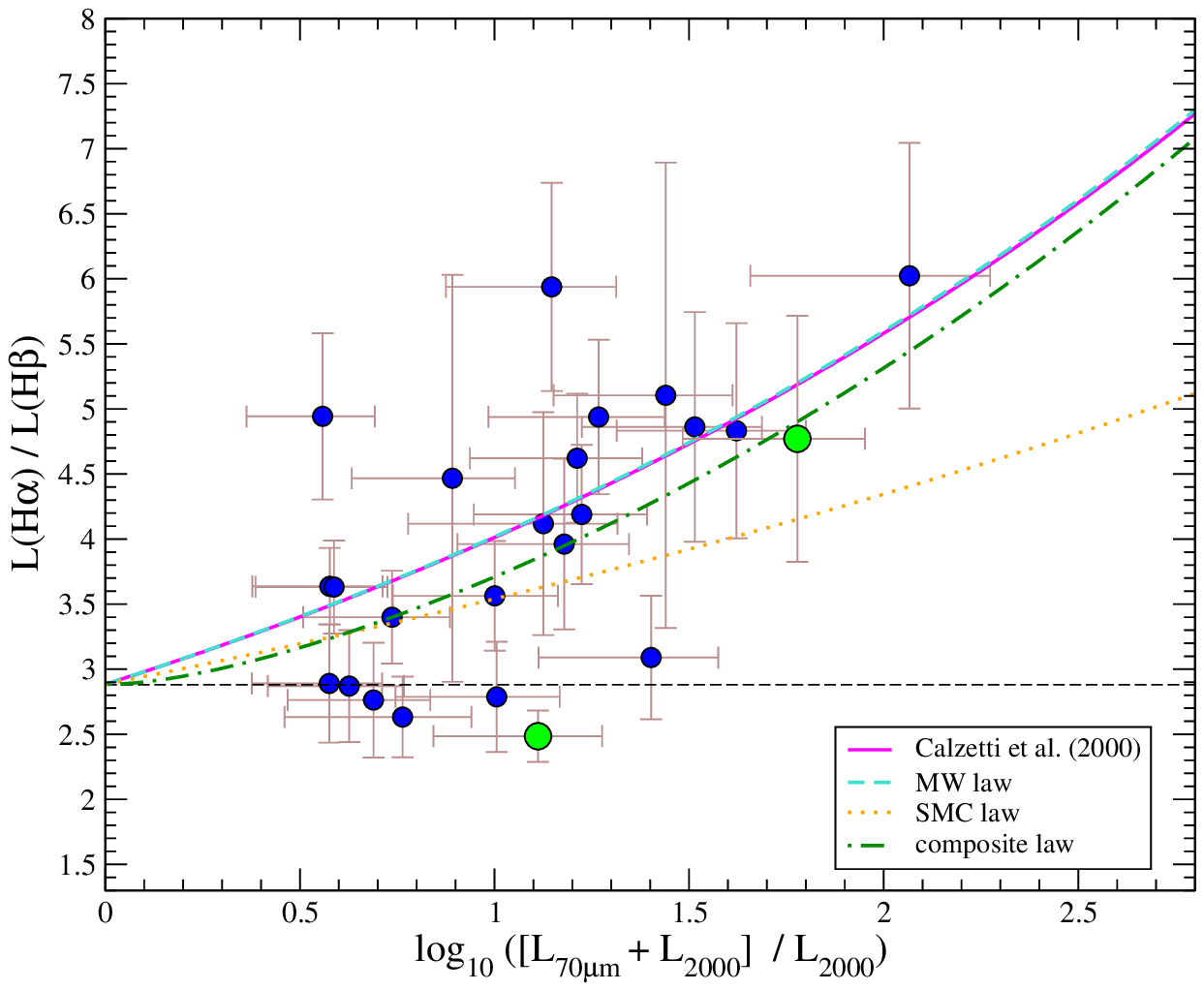}
\caption[]{\label{fig_redd} Balmer decrement between the \ha and \hb lines for our \tfm galaxies at $0.2<z<0.3$, versus the logarithm of their (far-IR+UV) over UV luminosity ratio.  We consider the latter quantity as proportional to the extinction in the UV. Only galaxies with EW$>5 \, \rm \AA$ for both the \ha and the \hb lines are shown. The colour and size code for the circles is the same as in previous figures. Thick lines of different styles and colours indicate different reddenning laws, as detailed in the inset within the plot. The horizontal dashed line marks the value of the intrinsic Balmer decrement in a case B recombination with temperature $T=10,000 \rm K$.  We note that some galaxies have $L(\rm H \alpha) / L(\rm H \beta)$ ratios that are equal to or lower than the case B recombination, but at the same time have significantly large  $(L_{\rm 70 \, \mu m}+L_{2000})/L_{\rm 2000}$ ratios. This is probably due to aperture effects, i.e. the 1-arcsec wide VIMOS slit maps the unobscured central region of the galaxy, while the IR flux is produced by surrounding dust (see Figure \ref{fig_stamps}).}
\end{figure}

Figure \ref{fig_redd} shows the observed Balmer decrement $L($\ha$)/L($\hb$)$  of these galaxies, versus the logarithm of the ratio between the (far-IR+UV) and the UV luminosities.
The latter quantity is related to the extinction at UV wavelengths $A_{2000}$, i.e. $(L_{\rm 70 \, \mu m}+L_{2000})/L_{\rm 2000} \approx 10^{0.4\,A_{2000}}$ (assuming a uniform geometry for the dust distribution; see e.g. Buat et al.~2002). To estimate the far-IR contribution, we used the rest-frame $70 \, \rm \mu m$ luminosity, which we compute based on the rest-frame \tfm luminosities and assuming the \nLntfm-$\nu L_\nu^{\rm 70 \, \mu m} \,$ relation calibrated by Bavouzet et al.~(2008).

The different thick colour lines indicate the relations obtained from different reddening laws. We see that both the MW and the Calzetti et al.~(2000) reddening laws produce basically the same curve. This is because of the following: we assumed that the observed \ha-to-\hb luminosity ratio is given by

\begin{equation}
\frac{L(\rm H \alpha)}{L(\rm H \beta)}=2.87 \times 10^{-0.4 \, (A_{\rm H_\alpha} - A_{\rm H_\beta})},
\end{equation}
 
\noindent i.e. the intrinsic case B recombination line ratio attenuated by the corresponding difference of extinctions. The extinction $A_{\rm H_\alpha, \rm H_\beta}$ is proportional to the reddening $k(\rm H_\alpha, \rm H_\beta)$, and thus

\begin{equation}
A_{\rm H_\alpha} - A_{\rm H_\beta} = A_{2000} \times \frac{k(\rm H_\alpha)-k(\rm H_\beta)}{k(2000 \, \rm \AA)}.
\end{equation}

\noindent This means that the attenuation on $L(\rm H \alpha) / L(\rm H \beta)$ is given by the slope of  each reddening curve between the \ha-\hb and the 2000 $\rm \AA$  spectral regions.

As the MW reddening law is characterised by the presence of a graphite bump around $2175 \rm \AA$ that is absent in the Calzetti et al. law, it  might seem surprising  that both curves in figure \ref{fig_redd} are nearly coincident. This is due to the wavelengths we considered to study the attenuation: the slopes of the two reddening laws are basically the same between these precise wavelengths. If we had used a different UV wavelength for our analysis, the two curves would be slightly different.

Both the MW and the Calzetti et al. reddening laws appear to correctly describe the average extinction observed in our \tfm galaxies, within the error bars.  The SMC law, instead,  has a quite flatter slope and tends to underestimate the observed $L(\rm H \alpha) / L(\rm H \beta)$ ratio.  So, to deredden the spectral lines in our galaxies, we  decided to adopt the Calzetti et al. reddening law, but  the MW law should also be suitable for the \hb-to-\ha spectral region  (as, in practice, the region of the graphite bump is convolved with the NUV filter response function, so we are not sensitive to this structure).

We observe in figure  \ref{fig_redd} that six galaxies in our sample display $L(\rm H \alpha) / L(\rm H \beta)$ ratios that are equal to or lower than the case B recombination, but at the same time have significantly large  $(L_{\rm 70 \, \mu m}+L_{2000})/L_{\rm 2000}$ ratios.  Five out of six of these galaxies are secure one-to-one \tfm-zCOSMOS identifications, so this behaviour is not due to incorrect associations. The $L(\rm H \alpha) / L(\rm H \beta)$ ratios  of these sources can neither be explained by a linear combination of two components, one completely free of dust (in a case B recombination) and the other one following e.g. the Calzetti et al. reddening law (dot-dashed green line in figure  \ref{fig_redd}).

Instead, the behaviour of the six outlier sources is probably the combination of two factors: 1) at $0.2<z<0.3$, the 1-arcsec wide VIMOS slits only cover a small region within the galaxies and 2) the dust is not homogeneously distributed; in particular, the optical spectra map  regions where the UV/optical photons can escape with virtually no absorption.

\begin{figure}
\plotone{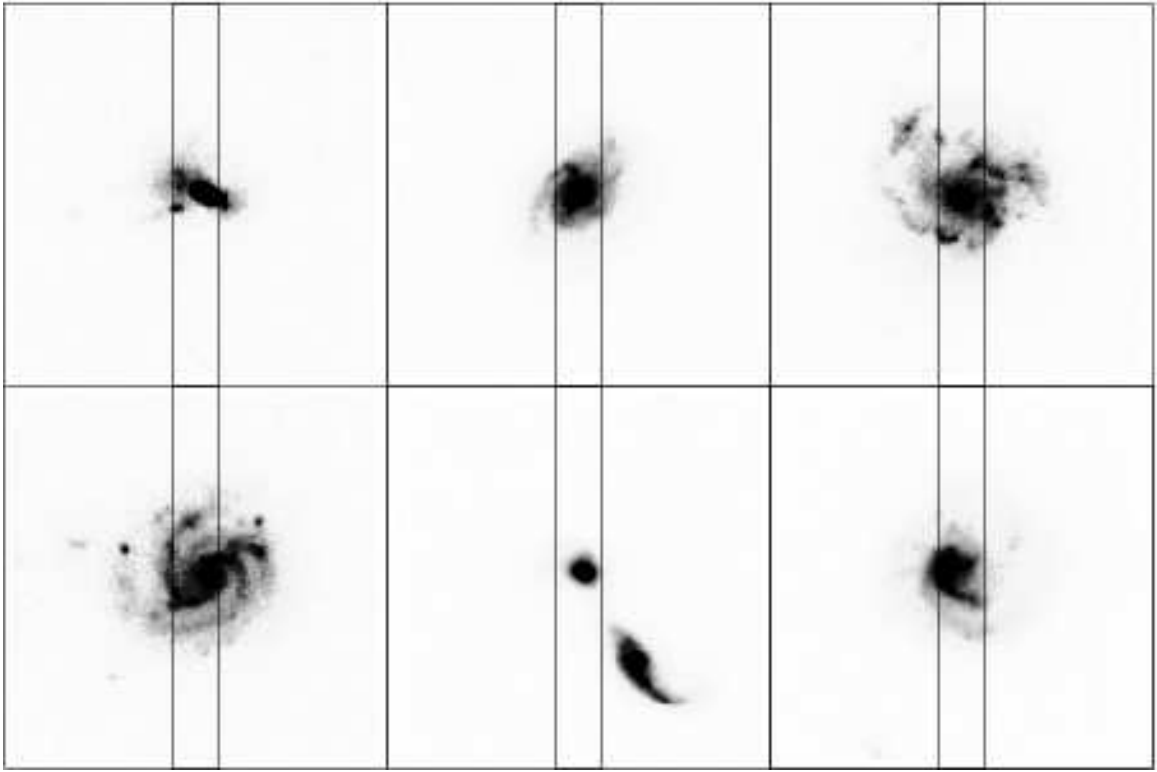}
\caption[]{\label{fig_stamps} {\em HST/ACS} $I$-band postage stamps of six galaxies characterised by optical spectra with an \ha-to-\hb ratio equal to or lower than the intrinsic absorption in the case B recombination. The size of each stamp is 7.5 $\times$ 7.5 arcsec. The position of the VIMOS slit is shown on each image.}
\end{figure}

The {\em HST/ACS} $I$-band images show that all five cases with secure identifications are galaxies with a bright optical nucleus and a quite fainter disky structure around (see figure \ref{fig_stamps}). This fainter structure is probably extincted by dust, which is itself responsible for the IR emission. The remaining sixth case corresponds to a merger, and the IR emission is likely to be produced by the dust surrounding the two merging galaxies.

Except for these few cases with resolved internal regions, the application of a single reddening law should be suitable for the majority of our galaxies. Although sources with non-homogeneous dust distributions might also be present at higher redshifts, the possibility of resolving different regions within a single galaxy is much less likely.

\begin{figure}
\epsscale{0.60}
\plotone{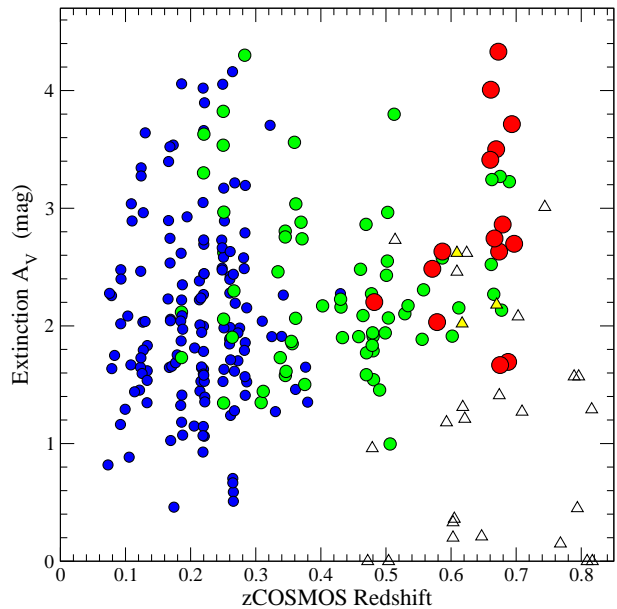}
\caption[]{\label{fig_av}  $V$-band extinctions  derived for the \tfm galaxies versus redshift, using the Calzetti et al. reddening law. In each case, we obtained the extinction by equating the SFR derived from the \ha or \hb lines with the total (IR+UV) SFR.  The colour and size code for the circles is the same as in previous figures. Triangles  indicate galaxies from the CFRS (Lilly et al.~1995), for which SFR and extinctions have been computed by Maier et al.~(2005). For the three CFRS galaxies that have been detected at 15 $\rm \mu m$ with {\em ISO} (Flores et al.~1999; filled triangles), we recomputed the SFR and extinctions taking into account their IR flux.}
\end{figure}

Figure \ref{fig_av} shows the  $V$-band extinctions derived  for different IR luminosity galaxies at different redshifts in our sample, assuming the validity of the Calzetti et al.~(2000) reddening law. To obtain the extinction, we imposed the SFR derived from the \ha or \hb line flux to be equal to the total (IR+UV) SFR in each galaxy.  As before, zCOSMOS BLAGN are excluded from this analysis.

We observe an evolution of the median extinction values as a function of IR luminosity and redshift. The \nLnn galaxies within our sample have  a median extinction $A_{V}=(1.98 \pm 0.73)$ mag, at a median redshift $z=0.22$. Among the \nLnu galaxies,  the median extinction is $A_{V}=(2.70 \pm 0.75)$ mag, at a median redshift $z=0.67$. These values are consistent with previous spectroscopic studies of IR galaxies (e.g. Liang et al.~2004; Choi et al.~2006). 

For a comparison, we included in figure \ref{fig_av} the extinctions obtained by Maier et al. (2005) from the spectra of 30 Canada-France Redshift Survey (CFRS; Lilly et al.~1995) sources at $0.5<z<0.9$. We see that most of the CFRS sources have smaller extinctions than our galaxies at similar redshifts. This is expected, as only a few of these 30 CFRS sources are IR luminous (Flores et al.~1999).

Choi et al. found that the extinction values within their IR galaxy sample were mainly correlated with IR luminosity rather than redshift. To test this property on our \tfm galaxies, we computed the correlation coefficients between extinction and IR luminosity, and between extinction and redshift:

\begin{equation}
\rho \, (A_{V}; y)=\frac{1}{N} \sum_{i=1}^N \frac{(A_{V}^i - \overline{A_{V}}) (y^i - \overline{y})}{\sigma_{A_{V}} \sigma_y},
\end{equation}

\noindent where the variable $y$ alternatively indicates rest-frame \tfm luminosity \nLntfm or the redshift $z$. The symbols with an overline indicate mean values and the different $\sigma$ correspond to  one-variable dispersions. The sum is made over all the galaxies considered to probe the correlations.

We firstly restrict this measurement to a given luminosity bin (\nLnl) and study the correlation between $A_{V}$ and $z$. We find  that the correlation coefficient  is $\rho \, (A_{V}; z) \approx -0.08$ (for $N=67$ galaxies), i.e. there is basically no correlation between the two variables $A_{V}$ and $z$.

Instead, we can fix the redshift bin and study the correlation between $A_{V}$ and \nLntfm. For the 93 galaxies at $0.2<z<0.45$, we find $\rho \, (A_{V}; \nu L_\nu^{\rm 24 \, \mu m}) \approx 0.07$. For the 48 galaxies at $0.45<z<0.70$, we have $\rho \, (A_{V}; \nu L_\nu^{\rm 24 \, \mu m}) \approx 0.46$. Thus,  within our sample, $A_{V}$ appears to be  somewhat correlated with IR luminosity for the intermediate redshift $0.45<z<0.70$ galaxies (at redshifts comparable to those explored by Choi et al.).

\section{Line diagnostics and metallicities}
\label{sec_line}

\subsection{\oiii/ \hb}
\label{sec_oiiihb}

We further explored the optical spectral characteristics of the \tfm galaxies in COSMOS by analysing the variation of typical line ratios as a function of IR luminosity and redshift. Whenever possible, we used these line ratios to estimate galaxy metallicities. However, in contrast to metal abundances, direct line ratios have the advantage of being quite model-independent  and free of physical assumptions.

\begin{figure}
\epsscale{0.9}
\plotone{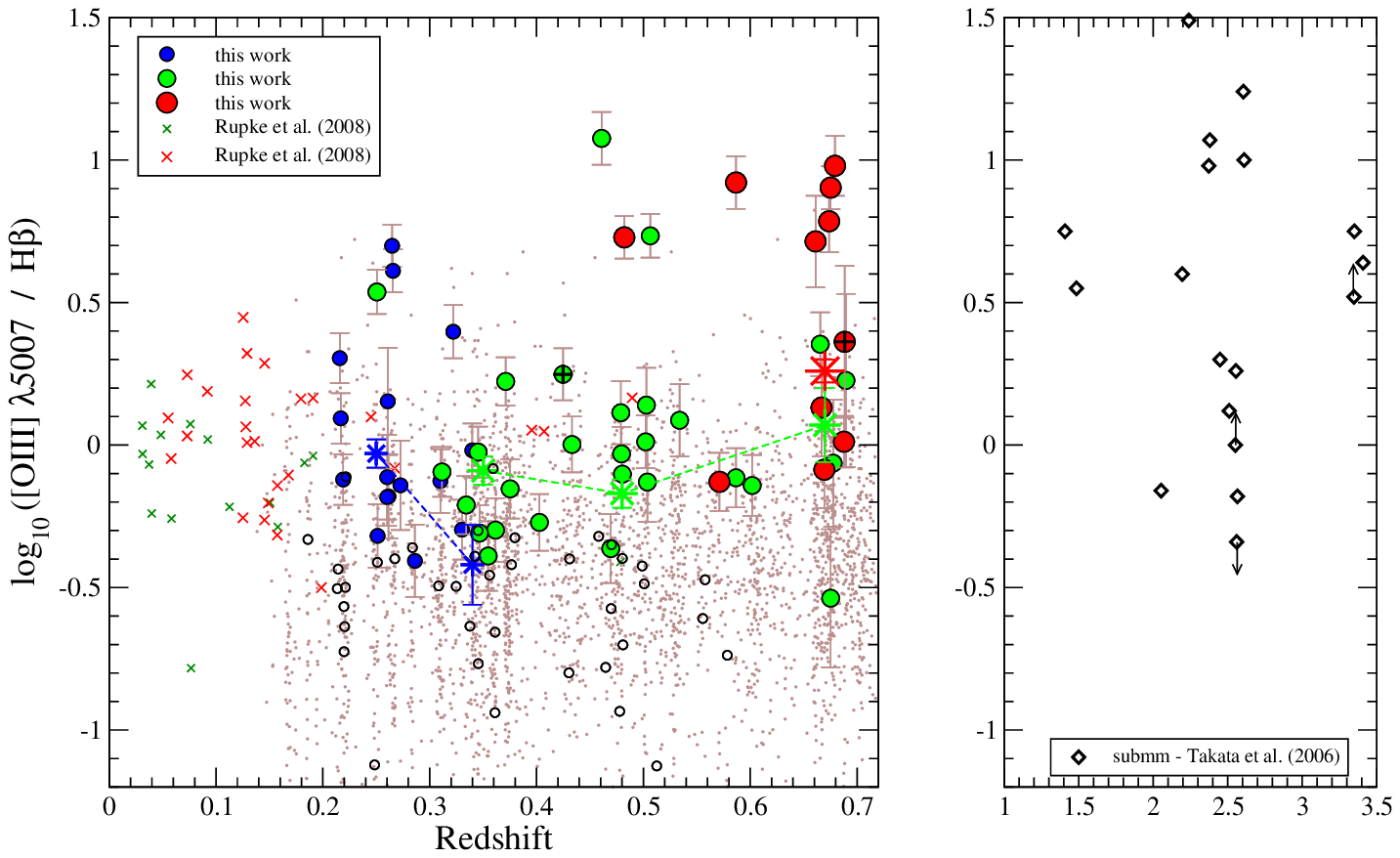}
\caption[]{\label{fig_hboiii} Left panel:  \oiii/\hb ratios for our galaxies in the cases that both \hb and \oiii have $\rm EW \geq 5 \, \rm \AA$, versus redshift. The colour and size code for  circles is the same as in previous figures. Filled circles with a plus sign within correspond to (narrow-line) X-ray AGN with $L_{X}>10^{42.5} \, \rm erg \, s^{-1}$. Measurements for those cases with \hb $\rm EW \geq 5 \, \rm \AA$ but \oiii $\rm EW < 5 \, \rm \AA$ are indicated with  small empty circles.  Asterisks correspond to our measurements on average stacked spectra. The colour and size code for the asterisks is equivalent to that for circles. Brown dots correspond to measurements performed on all other zCOSMOS galaxies with \hb EW $> 5 \, \rm \AA$ at similar redshifts (Lamareille et al., in preparation). Small green and large red crosses correspond to measurements listed in Table 2 of Rupke et al.~(2008) for LIRGs and ULIRGs, respectively. Right panel: \oiii/\hb ratios for sub-millimetre galaxies at $z\gsim 1.5$ obtained by Takata et al.~(2006).}
\end{figure}

Given the wavelength range covered by the zCOSMOS spectra, \hb and \oiii can be measured for galaxies at $0.2<z<0.7$ without being significantly affected by fringing. The left panel of figure \ref{fig_hboiii} shows the \oiii/\hb ratios as a function of redshift. Filled circles  correspond to our measurements on the individual spectra, in the cases when the two line EW were  $\geq 5 \, \rm \AA$. The size and colour code for the circles is the same as in previous plots. Small empty circles indicate tentative \oiii/\hb measurements on spectra with 
\hb EW $\geq 5 \, \rm \AA$ but \oiii EW $< 5 \, \rm \AA$. The zCOSMOS BLAGN are excluded from this analysis. A single case of the shown line ratios corresponds to an obvious  narrow-line AGN, given its high X-ray luminosity $L_{X}>10^{42.5} \, \rm erg \, s^{-1}$ (filled circle with a plus sign within). All line measurements include aperture and extinction corrections, although these corrections have little impact on the derived ratios. \hb is also corrected for stellar absorption in all cases. All the completely corrected flux measurements in the cases that both lines have EW $\geq 5 \, \rm \AA$ are listed in table \ref{tab_hboiii}. 

To study the average line ratios of all our galaxies independently of the selection effects imposed by the 5 $\rm \AA$ EW cut, we measured the line fluxes on  average composite spectra in different IR luminosity and redshift bins (asterisks in figure \ref{fig_hboiii}). To construct each composite, we considered all the galaxies in each bin, independently of the line EW measured on their individual spectra. We assigned to each composite a redshift equal to the median of the individual galaxy redshifts. 

Also, for a comparison, we show the \oiii/\hb ratios of all the zCOSMOS galaxies with \hb EW $\geq 5 \, \rm \AA$ (Lamareille et al., in preparation; brown dots in figure \ref{fig_hboiii}). The average of these ratios are remarkably constant with redshift, with $\log_{10}($\oiii$/$\hb$) \approx -0.5$ at $z=0.2-0.3$ and $\log_{10}($\oiii$/$\hb$) \approx -0.4$ at $z=0.6-0.7$.

By comparison of the \oiii/\hb line ratios among our galaxies with different luminosities and redshifts,  and with the line ratios of the entire zCOSMOS galaxy population, we can conclude the following:

\begin{itemize}
\item the average \oiii/\hb values obtained from the stacked spectra for \nLnn galaxies at $z\sim 0.25$ are comparable to those of \nLnl sources at $z\sim 0.4-0.7$. This indicates that the hardness of the ionising flux is comparable in all these galaxies. This fact suggests that the fraction of massive stars among all the stars being created in these galaxies  are similar. 

\item For the less luminous IR galaxies in our sample at $0.30<z<0.45$, the average ionisation level is quite low and comparable to the average of the entire zCOSMOS sample. We obtained the average of our IR galaxies by stacking 38 galaxies with \nLnn at redshifts $0.30<z\leq 0.45$. This fact indicates that the ionising fluxes of these sources of moderate IR luminosity are similar to those of any other optically bright galaxy at low redshift.

\item the average \oiii/\hb ratio obtained from the composite spectrum of 42 galaxies with \nLnu  at a median redshift of $z=0.67$ is more than 0.6 dex higher than the average ratio of all the zCOSMOS galaxies at similar redshifts.  This result shows that the most luminous IR galaxies at $z=0.6-0.7$ have different chemical properties from other more typical optically-selected sources. In particular, among our \nLnu sources at $z>0.45$   with  \hb and \oiii EW $> 5\, \rm \AA$ ,  more than a half have $\log_{10}($\oiii$/$\hb$) > 0.5$. These very high ionisation levels --characterising only $\sim$1\% of the total zCOSMOS galaxies at these redshifts-- could suggest the presence of some remaining (narrow-line) AGN within our sample, but could also indicate that a large fraction of young massive stars of types O and B are being created in these galaxies.  
\end{itemize}

  We also compared our derived line ratios with other measurements available in the literature. Small green and large red crosses in the left-hand panel of figure \ref{fig_hboiii} show the \oiii/\hb line ratios obtained by Rupke et al.~(2008) on a sample of {\em IRAS}-selected LIRGs and ULIRGs, respectively, mostly at $z<0.3$. None of the sources in Rupke et al. have the high ionisation levels characteristic of several of our \nLnu galaxies at $0.5<z<0.7$ and a few of the \nLnl galaxies at lower redshifts (although Veilleux et al.~1995 did find a few cases of ULIRGs with high ionisation levels in the local Universe).
  
  The high ionisation levels characterising some of our \nLnu sources at $0.5<z<0.7$ are comparable to those found in many sub-millimetre galaxies at $z \gsim 1.5$ (Takata et al.~2006; right-hand panel of figure \ref{fig_hboiii}). Unfortunately, the determination of the source of ionisation in both these sub-millimetre galaxies and our intermediate-redshift ULIRGs requires much better line sensitivity than that provided by low-to-medium resolution spectra. Future follow up of these sources with near or mid-IR spectroscopy should help us to better constrain the nature of our  most luminous IR galaxies at intermediate redshifts.

\subsection{\nii/\ha and the BPT diagram}

\begin{figure}
\epsscale{0.60}
\plotone{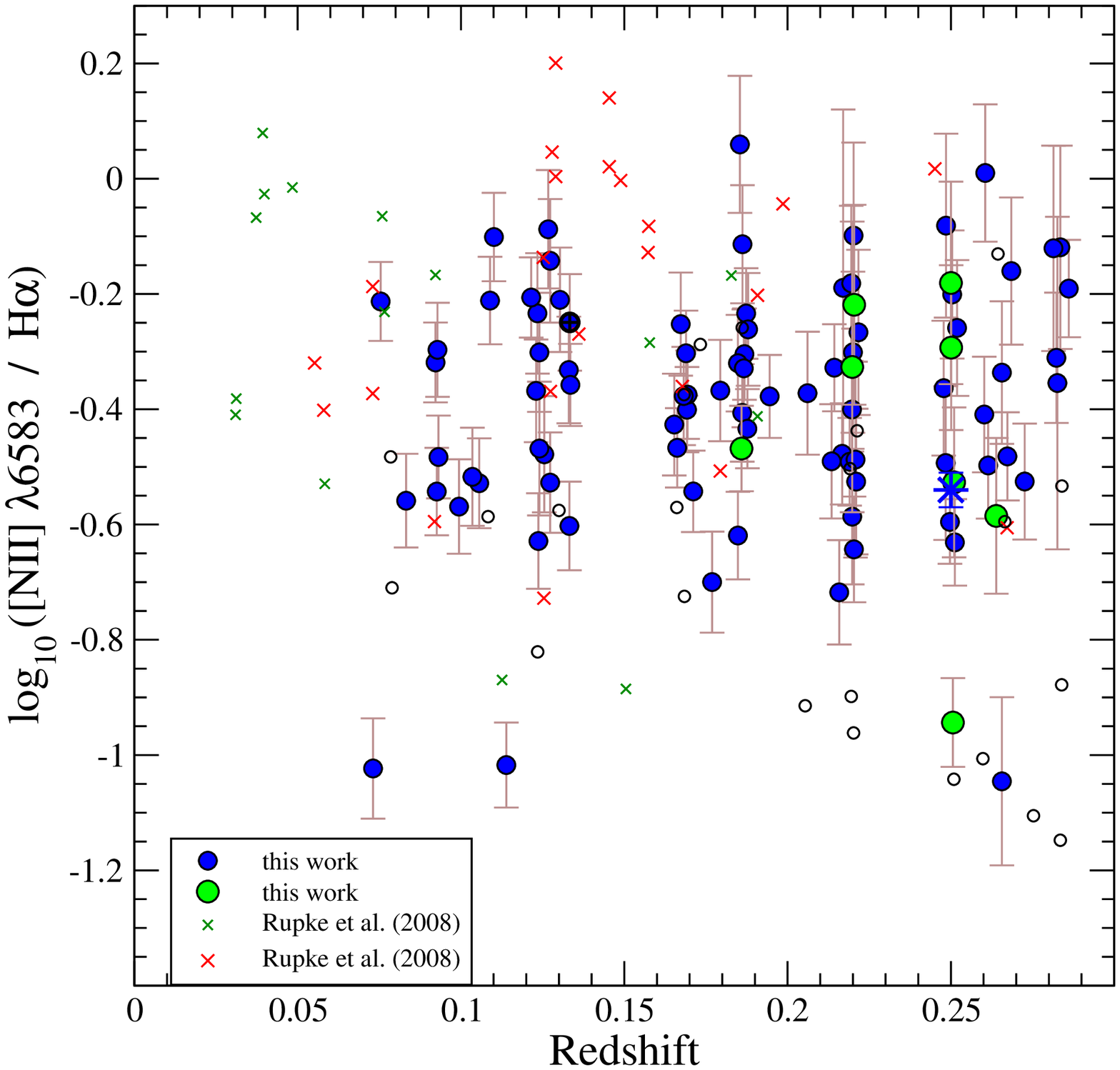}
\caption[]{\label{fig_hanii}  \nii/\ha line ratios versus redshift for our galaxies in the cases that both \ha and \nii have $\rm EW > 5 \, \rm \AA$. The colour and size code for circles is the same as in previous figures. Small empty circles indicate measurements for sources with \ha $\rm EW \geq 5 \, \rm \AA$ but \nii $\rm EW < 5 \, \rm \AA$.}
\end{figure}

   For galaxies up to redshift $z\sim0.3$, the \ha and \nii lines are observed in the zCOSMOS spectra and quite unaffected by fringing. The completely-corrected \ha and \nii fluxes for our \tfm galaxies with EW $>5 \, \rm \AA$ are presented in table \ref{tab_hanii} and the corresponding \nii/\ha ratios are shown in figure \ref{fig_hanii}. We see that our \tfm galaxies span a wide range of \nii/\ha ratios. This is in part due to the same spatial sampling effect discussed in Section \ref{sec_spec}, which produced the large dispersion in the \ha EW. 
  
\begin{figure}
\epsscale{0.70}
\plotone{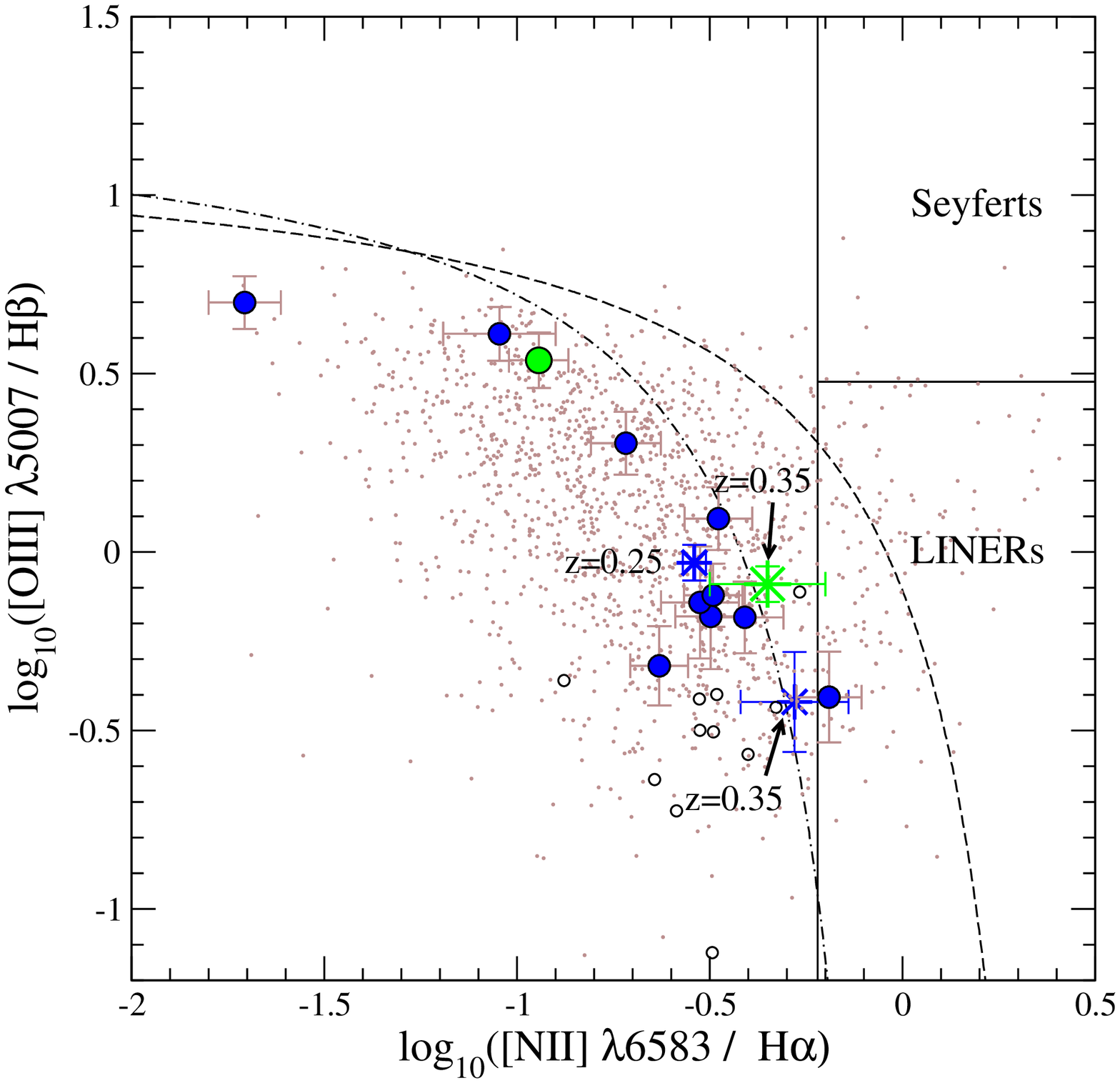}
\caption[]{\label{fig_bpt} The BPT diagram for our \tfm sources at $0.2<z<0.3$ with $\rm EW > 5 \, \rm \AA$ for the four \hb, \oiii, \ha and \nii lines. Small empty circles indicate  sources for which only \ha and \hb have $\rm EW > 5 \, \rm \AA$. Asterisks indicate measurements on composite spectra at median redshifts $z=0.25$ and $0.35$. The colour and size code for circles and asterisks is the same as in previous figures.  Brown dots correspond to measurements performed on all other zCOSMOS galaxies with \ha and \hb EW $> 5 \, \rm \AA$ at similar redshifts (Lamareille et al., in preparation). The dashed line indicates the starburst/AGN division line proposed by Kewley et al.~(2001), while the dot-dashed line corresponds to the limit for composite systems (Kauffmann et al.~2003a).}
\end{figure}

  For $\sim 15\%$ of our \tfm galaxies at $z<0.3$, the  ratios are quite high  $\log_{10}($\nii/ \ha$)>-0.2$. Although we do not have reliable \hb and \oiii measurements for most of them, the high \nii/ \ha ratios alone are sufficient to classify these objects either as Seyferts or as their low-luminosity equivalents, i.e low-ionisation nuclear emission-line regions (LINERs; see figure \ref{fig_bpt}). These high \nii/ \ha values are more common among more luminous IR galaxies at similar redshifts (Rupke et al.~2008; crosses in figure \ref{fig_hanii}).

   Our \tfm galaxies at $0.2<z<0.3$ with  measurements for the four  \hb, \oiii, \ha and \nii lines can be located  in the Baldwin, Phillips \& Terlevich  (1981; BPT) diagram (figure \ref{fig_bpt}). Once more, we compare our measurements with the line ratios of other zCOSMOS galaxies at similar redshifts (Lamareille et al., in preparation).  Nearly all of our  \tfm galaxies at these redshifts have  mid-IR luminosities \nLnn and occupy quite different positions in the BPT diagram, similarly to other optically-selected galaxies that are below our detection limits in the IR. This indicates the quite heterogeneous character of the IR normal galaxy population: any galaxy with modest levels of SF ($\lsim 10-15\, \rm M_\odot \, yr^{-1}$) belongs to this category, independently of its degree of chemical evolution and already-assembled stellar mass.

\begin{figure}
\epsscale{0.8}
\plotone{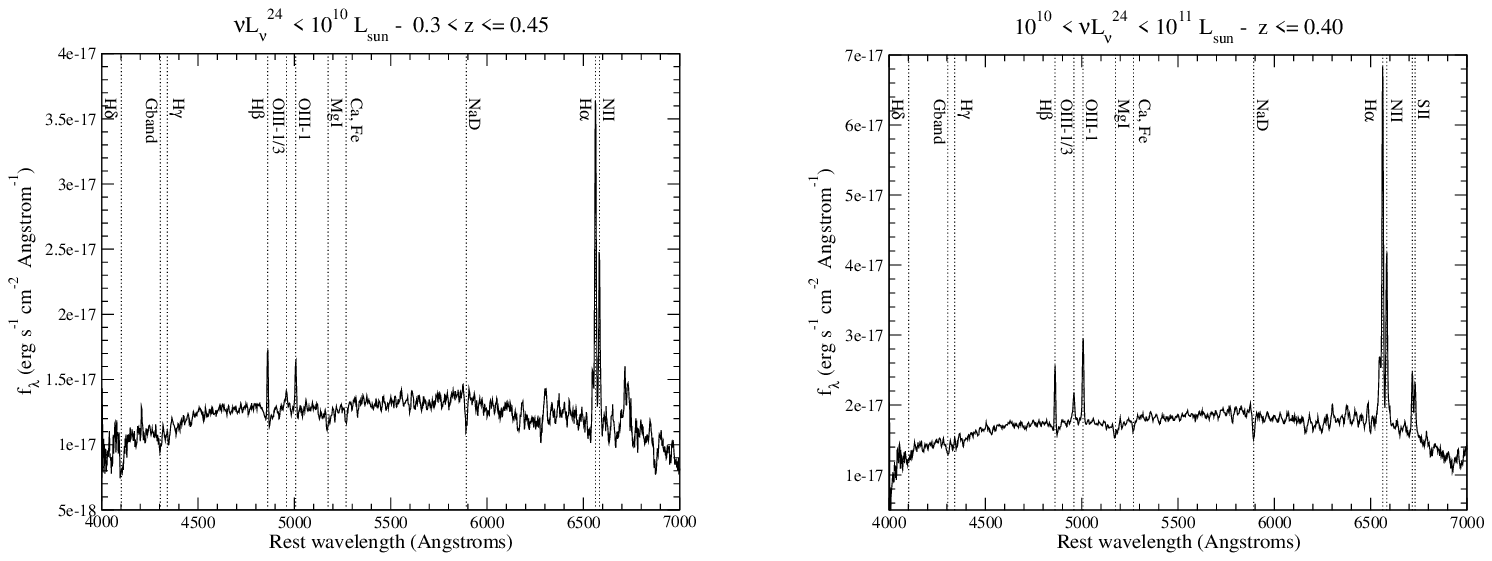}
\caption[]{\label{fig_stackfring} The same composite spectra shown in the bottom panel of figure \ref{fig_stack1} and in the top panel of figure \ref{fig_stack2}, but in this case extended beyond observed wavelength $\lambda=8500 \, \rm \AA$, i.e. the region significantly affected by fringing. The composite has been made by stacking the spectra of 38 and 70 galaxies, respectively,  with \nLnn and \nLnl, and with median redshifts $z=0.34$ and $0.35$.}
\end{figure}

 The asterisks in figure \ref{fig_bpt} indicate our lines ratios measured on the average stacked spectra at median redshifts $z=0.25$ and $\sim 0.35$. Although we did not measure \ha and \nii line fluxes at $z>0.3$  on an individual basis to avoid being severely affected by fringing, the signal-to-noise ratio in the stacked spectra  is sufficiently good as to allow for a reliable measurement of these lines (see figure \ref{fig_stackfring}). These average spectra correspond to \nLnl and \nLnn galaxies and their line ratios  are located just beyond the empirical line proposed by Kauffmann et al.~(2003a), which delimits the transition region between pure star-forming and mixed AGN/star-forming galaxies in the BPT diagram. This fact suggests that several mid-IR-selected galaxies at $z\sim0.3-0.4$ redshifts, even with modest IR luminosities,  might have a composite star-forming/AGN nature.

\subsection{Oxygen abundances}  
\label{sec_met}

 We used the measured and corrected line fluxes  to estimate the oxygen abundances [O/H] of our  \tfm-selected  galaxies, whenever possible. 
 
  Several methods exist to derive abundances from emission-line ratios, all of which have both advantages and disadvantages. For example, the Pettini \& Pagel~(2004) formulation uses both the \oiii/\hb and \nii/\ha ratios, but the \nii/\ha ratio is particularly affected by AGN radiation fields, so the resulting abundance can be very uncertain when part of the spectral light has a non-stellar origin. The more widely used $R_{23}$ parameter (Pagel et al.~1979) is based on the \oii, \hb and $\rm [OIII] \, \lambda\lambda 4959,5007 \,$ lines, all of which are usually observed in optical spectra. However, the oxygen abundance is two-valued with respect to $R_{23}$, and any given value of $R_{23}$ lower than the maximum is compatible with a low- and a high-metallicity solutions. When available, additional emission lines are sometimes used to try to lift this degeneracy. Another common approach is to consider only one of the branches (usually that of high-metallicity) in the [O/H] versus $R_{23}$ relation.  Further discussion about different abundance estimators can be found in e.g. Kobulnicky et al.~(1999) and  Kewley \& Dopita~(2002).
 
 We used the algorithm described by Maier et al.~(2005) to derive oxygen abundances, which is based on the Kewley \& Dopita~(2002) model.  This algorithm also uses the information on different line fluxes, but performs a simultaneous fit to all available emission lines, with three free parameters: the extinction, the ionisation parameter (which is proportional to the flux of ionising photons) and the abundance [O/H].
 
 We implemented this algorithm in two redshift ranges: 1) $0.2<z<0.3$, for the same galaxies shown in the BPT diagram, for which we have \hb, \oiii, \ha and \nii  measurements; 2) $0.5<z<0.7$, for galaxies with  \oii, \hb and \oiii measurements. In all cases, we fixed the value for the extinction to that derived from the balance between the \ha or \hb and the total (IR+UV) SFR, i.e. we used the algorithm with only two free parameters. We excluded from this analysis a few sources belonging to the group of galaxies with \hb/\ha ratios close to a case B recombination, but large $(L_{\rm 70 \, \mu m}+L_{2000})/L_{\rm 2000}$  ratios (see Section \ref{sec_dust}). All zCOSMOS BLAGN are also excluded (as they were from all our previous line diagnostic).

  For sources in the redshift range 1), the algorithm nearly always produces either non-degenerate solutions or two solutions of  relatively high-abundance, in such a way that a useful lower limit can be set. Galaxies in range 2), in contrast, have the same problem as when using the $R_{23}$ parameter: unless we are close to a maximum $R_{23}$ value, there are one possible high- and one low-metallicity solutions. 

\begin{figure}
\epsscale{0.65}
\plotone{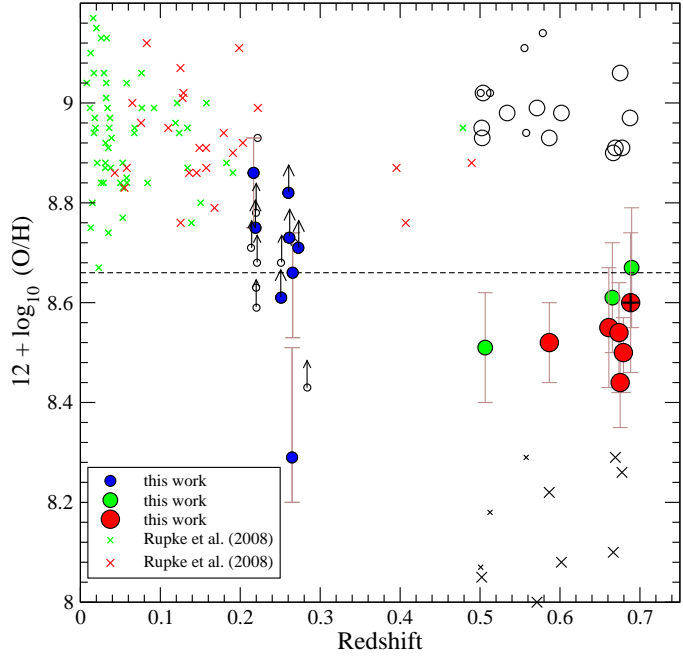}
\caption[]{\label{fig_met} Oxygen abundances derived for \tfm/zCOSMOS galaxies at different redshifts. All the abundances have been calculated with the same technique, assuming the dust extinction to be known and fixed (see text). Cases of non-degenerate abundances are indicated by filled circles with errors bars.   Upward-pointing arrow indicate lower limits.  Cases of degenerate metallicities are represented by empty circles (upper branch) and crosses (for the corresponding lower-branch values). The colour and size code for circles is the same as in previous figures. The circle with a plus sign within corresponds to a (narrow-line) X-ray AGN with $L_{X}>10^{42.5} \, \rm erg \, s^{-1}$. The dashed line indicates the value of the solar metallicity determined by Asplund et al.~(2004).}
\end{figure}

Figure \ref{fig_met} shows the obtained abundance values as a function of redshift. Non-degenerate cases are indicated with filled circles with error bars. Circles with upward-pointing arrows at $z<0.3$ indicate lower limits. As before, small empty circles represent galaxies for which \ha and \hb EW $> 5 \, \rm \AA$, but with at least one of the other line EW below that cut.

Among our 26 galaxies with computed abundances at $0.5<z<0.7$, 9 ($\sim$35\%) have non-degenerate values.  For the remaining 17, there is an upper-branch (large empty circles) and a lower-branch (large black crosses) solutions.  The dashed line indicates the solar abundance derived by Asplund et al.~(2004). As we have computed all the abundances using the same technique, it makes sense to compare the results we obtained at different redshifts.

The $\sim 35\%$ of non-degenerate abundances at redshifts $0.5<z<0.7$ are  solar or sub-solar, while most of the lower redshift metallicities at $0.2<z<0.3$ are solar or super-solar. We note that the important point here is the relative values of the abundances at different redshifts. The fact they are sub- or super-solar is incidental in our comparison, as actually different metallicity indicators have systematic differences of up to $\sim 0.5 \,\rm dex$ among them (e.g. Kewley \& Dopita~2002; Ellison \& Kewley~2006). For example, if we used the Pettini \& Pagel~(2004) prescription for our galaxies at $0.2<z<0.3$, we would obtain abundances which are $\sim 0.2-0.3 \, \rm dex$ lower than those shown in figure \ref{fig_met}.

\begin{figure}
\epsscale{0.65}
\plotone{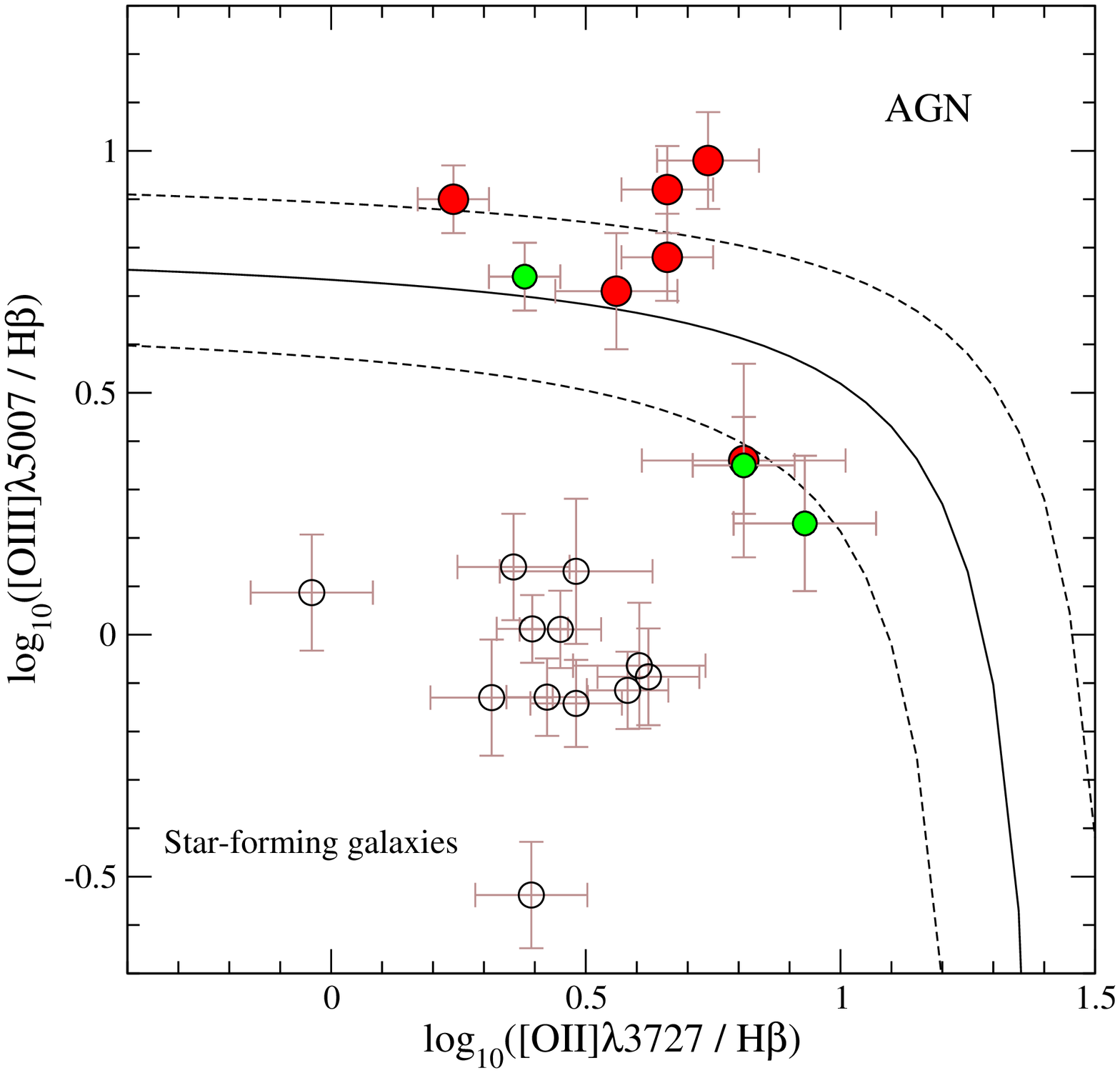}
\caption[]{\label{fig_oiioiii}  \oiii/\hb versus \oii/\hb ratios for the galaxies with derived  metallicities at $0.5<z<0.7$ shown in figures \ref{fig_met} and \ref{fig_metmass}. Filled circles indicate cases of non-degenerate metallicities, while empty circles correspond to cases with degenerate values. The solid and dashed lines are, respectively, the median and confidence limits of the empirical separation between star-forming galaxies and AGN derived by Lamareille et al.~(2004). The (narrow-line) X-ray AGN with $L_{X}>10^{42.5} \, \rm erg \, s^{-1}$ has $\rm log_{10}([OII/H\beta])=0.81\pm0.20$ and $\rm log_{10}([OIII/H\beta])=0.36\pm0.20$ (red large circle behind the green small one).}
\end{figure}

At $0.5<z<0.7$, non-degenerate abundances correspond to cases with large $R_{23}$ values, i.e. with relatively high  \oiii/\hb  and \oii/\hb ratios. These ratios are illustrated in figure \ref{fig_oiioiii}. We see that all these galaxies with non-degenerate abundances  lie within or very close to the star-forming galaxies/AGN separation region proposed by Lamareille et al.~(2004). This is a potential concern in our analysis: the derived metallicities of these objects could be affected by the plausible presence of an AGN.  We note, however, that there is only one  X-ray AGN among these galaxies and it lies just below the separation region (red large circle behind the green small one), showing that the association of high ionisation lines and AGN activity is not straightforward and has to be taken with care. Besides, strong ionisation lines such as \oii and \oiii --resulting in very high \oiii/\hb and \oii/\hb values-- can also be produced by very young massive stars (see e.g. Charlot \& Longhetti~2001), with no need of invoking the presence of an AGN.

The galaxies with degenerate abundances at $0.5<z<0.7$ are likely to have high metallicities (see the mass-metallicity relation discussion in Section \ref{sec_mm} and figure \ref{fig_metmass}). They constitute $\sim$65\% of our higher-redshift metallicity sample. These upper-branch metallicity values are comparable to those we obtained at lower-redshift sources and those derived for {\em IRAS}-selected LIRGs and ULIRGs by Rupke et al.~(2008;  small green and red crosses in figure \ref{fig_met}). The remaining 35\% of our galaxies at $0.5<z<0.7$ have lower abundances than the vast majority of low-redshift IR sources. These results suggest that at least $1/3$ of the hosts of the most intense IR emission have chemically evolved from redshifts $z\sim0.7$ to $z\sim0.2$.

\subsection{The mass-metallicity relation}
\label{sec_mm}

 The mass (or luminosity)-metallicity relation for different galaxy populations has been the subject of multiple studies in the literature (e.g. Zaritsky et al.~1994; Contini et al.~2002;  Lilly et al.~2003;  Tremonti et al.~2004), and it has been found to evolve with redshift (e.g. Kobulnicky et al.~2003; Contini et al.~2008). This relation can be explained within the context of hierarchical models of galaxy formation, and by including the effects of supernovae-driven winds which help to remove metals especially out of the less-massive galaxies  (e.g. Tissera et al.~2005; Brooks et al.~2007).

\begin{figure}
\epsscale{0.5}
\plotone{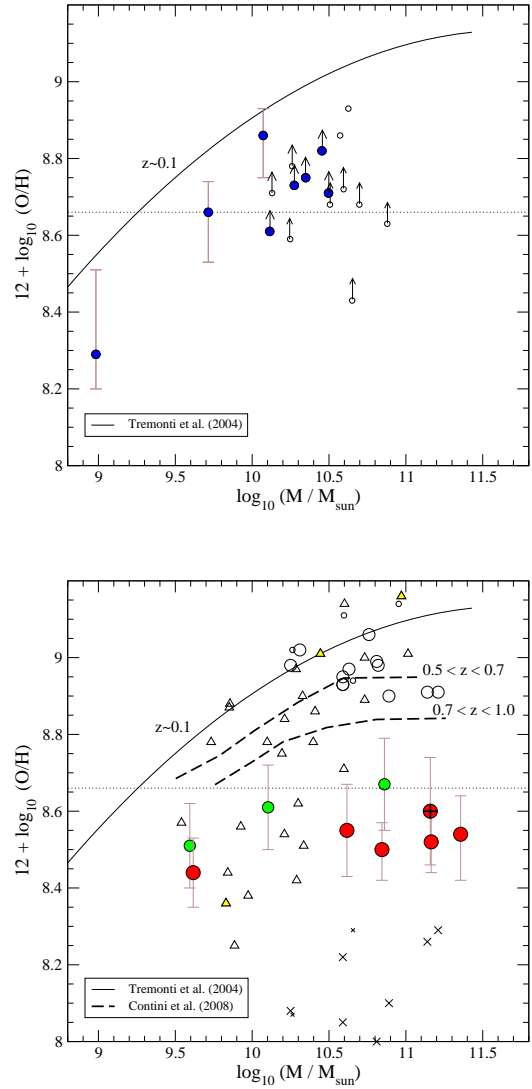}
\caption[]{\label{fig_metmass} Oxygen abundances versus already-assembled stellar masses for our \tfm-selected galaxies. Top panel: $0.2<z<0.3$ galaxies. Bottom panel: $0.5<z<0.7$ galaxies.
The symbol code is the same as in previous figures. Triangles in the bottom panel indicate galaxies from the CFRS (Lilly et al.~1995) with known metallicities (Maier et al.~2005). Those triangles filled in yellow correspond to CFRS galaxies with {\em ISO} 15 $\rm \mu m$ detections (Flores et al.~1999). The solid line shows the best-fit mass-metallicity relation obtained by Tremonti et al.~(2004) on SDSS data. The dashed lines correspond to its evolution at $0.5<z<0.7$ and $0.7<z<1.0$, as derived from all zCOSMOS-bright 10k galaxies (Contini et al.~2008).}
\end{figure}

 It is then of interest to see whether IR-selected galaxies follow a similar relation. Figure \ref{fig_metmass} shows metallicities versus already-assembled stellar masses for our \tfm galaxies (at $0.2<z<0.3$ and $0.5<z<0.7$, in the top and bottom panels, respectively). Stellar masses for all zCOSMOS galaxies have been computed by fitting Bruzual \& Charlot models to the multiwavelength (U-band to 4.5 $\mu \rm m$) SED of these sources (Bolzonella et al., in preparation). All stellar masses assume a Salpeter (1955) initial mass function (IMF) over stellar masses $M=(0.1-100) \, M_\odot$. As before, filled  circles correspond to non-degenerate metallicity values, while empty circles and crosses show the upper and low-branch solutions, respectively, in all the degenerate cases.  For a comparison, we have also included  a sample of $0.5 \lsim z \lsim 0.9$ galaxies from the CFRS (triangles in the bottom panel of figure \ref{fig_metmass}). All these galaxies have secure metallicity determinations (Maier et al.~2005). Three of them have been detected by {\em ISO} at 15 $\rm \mu m$ (Flores et al.~1999; yellow filled triangles).

The solid line in both panels of figure \ref{fig_metmass}  is the best mass-metallicity relation fitted by Tremonti et al.~(2004) on Sloan Digital Sky Survey (SDSS) galaxies at $z\sim0.1$. The dashed lines in the bottom panel are the corresponding relations at $0.5<z<0.7$ and $0.7<z<1.0$, as obtained by Contini et al.~(2008) based on the entire zCOSMOS-bright 10k sample. Both the Tremonti et al. and the Contini et al. relations have been re-scaled to the Salpeter IMF.  Instead, we did not modify any of their metallicities.  The Tremonti et al. calibration produces very similar metallicities as the Kewley \& Dopita method (Ellison \& Kewley~2006). The Contini et al. metallicities are basically the same as ours for the common zCOSMOS objects. Thus, we left the Contini et al. metallicities in their original scale.

 Inspection of the top panel of  figure \ref{fig_metmass} shows that some of our  \tfm-selected galaxies at $0.2<z<0.3$ have metallicities which are at most marginally consistent with the general mass-metallicity relation at $z\sim0.1$. However, in many cases, we only show lower limits, so they still have the possibility of being  in agreement with the general mass-metallicity trend.
 
 At higher redshifts,  $\sim$65\% of our galaxies have one of the two possible metallicity values consistent with the general mass-metallicity relation derived by Contini et al.~(2008) for all zCOSMOS-bright 10k galaxies. However, most of our galaxies with a non-degenerate metallicity determination appear to have lower abundances  than expected. Only 2 out of these 9 secure-metallicity galaxies have stellar masses and abundances more or less in consistence with the extrapolation of the Contini et al.~ relations. The other 7 (i.e. $\sim 27\%$ of the total $0.5<z<0.7$ sample) are significantly out of the general mass-metallicity trend.

 Several of the CFRS galaxies shown in figure \ref{fig_metmass} have also been found to be under-abundant with respect to galaxies in the upper metallicity branch (Maier et al.~2005).  We note, however, that their characteristic stellar masses are, in general, quite smaller than for most of our galaxies. So, the discrepancies we observe for some of our IR galaxies are considerably more important.

As we discussed in Section \ref{sec_met}, most of these galaxies have high ionisation levels, so there exist the concern that the plausible presence of a narrow-line AGN  could be in part responsible for the low measured metallicities.    However, it is not clear that the discrepancies with the general mass-metallicity relation are due to the effects of a plausible AGN.  For example, one of the sources with the highest ionisation ratios $\rm log_{10}(OIII/H\beta)=0.90\pm0.10$ (see figure \ref{fig_oiioiii}) has a stellar mass consistent with the extrapolation of the Contini et al.~ relations. On the other hand, two out of three of the galaxies with secure metallicities that are just within the starburst region in the \oiii/\hb versus \oii/\hb diagram have metallicities significantly below the expected trend.

From the study of  a sample of {\em IRAS}-selected galaxies, Rupke et al.~(2008) found that the vast majority of LIRGs and ULIRGs at low $z<0.3$ redshifts are significantly under-abundant. We cannot confirm this from our metallicity measurements at similar redshifts, but the difference is probably due to the fact that our IR galaxies at $z<0.3$ are less IR-luminous than Rupke et al. galaxies and more similar to typical optical galaxies at low redshifts. 

Instead, we can confirm the under-abundance for a fraction ($\sim1/3$) of our \tfm sources at $0.5<z<0.7$. To explain the relatively low metallicities of their galaxies, Rupke et al. proposed that, during the IR phase,  less-abundant gas from outer parts of the galaxies could flow inward to the central regions and dilute the nuclear metal content. Although this mechanism appears to be common in gas-rich mergers (e.g. Barnes \& Hernquist 1996), its influence is less clear in regular star-forming disks, as many luminous IR galaxies at $z<1$ are known to be. Supernovae-driven outflows are also known to effectively remove metals from a galaxy (e.g. Dalcanton~2007) but it seems implausible that this mechanism alone can explain the under-abundance observed in IR galaxies (see the discussion in Rupke et al.~2008). More likely, a combination of the two phenomena could have taken place at the same time.  Thus, gas mixing produced by inflows and outflows during the IR phase could temporarily move galaxies appart from the mass-metallicity relation.

\section{AGN in the \tfm sample}
\label{sec_agn}

Optical spectra also allow us to test the presence of AGN within our \tfm sample. The zCOSMOS classification includes a special flag for  BLAGN, generally recognised by the presence of broad \ha or \hb lines, for the shape of the continuum, and also for the presence of MgII and CIII in emission in high-$z$ sources.

\begin{figure}
\epsscale{0.8}
\plotone{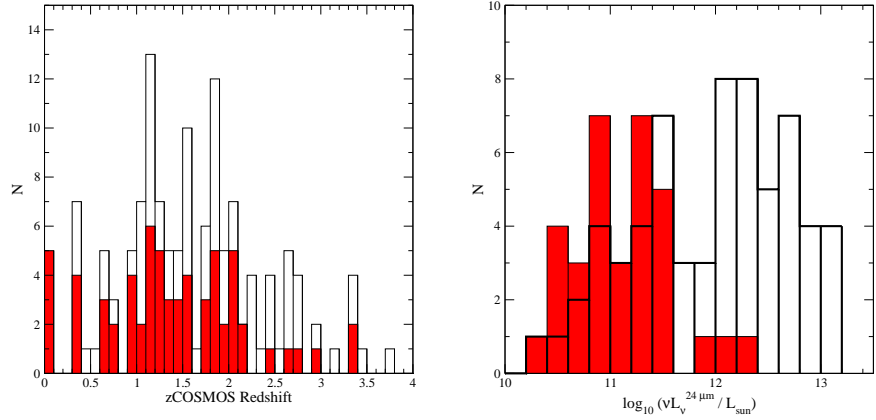}
\caption[]{\label{fig_agn} Left panel: redshift distributions of all BLAGN in the zCOSMOS-bright 10k catalogue (empty histogram) and those detected in the MIPS-shallow catalogue (filled histogram). Right panel:  \tfm luminosity distributions of all zCOSMOS-bright BLAGN in our sample (thick-line, empty histogram) and the narrow line IR sources associated with X-ray AGN with $L_{X}>10^{42.5} \, \rm erg \, s^{-1}$ (thin-line, filled histogram).}
\end{figure}

The entire zCOSMOS-bright 10k catalogue contains  132 BLAGN with good-quality spectroscopic redshifts and spectra, all of which have $I<22.5$ AB mag. 64 out of these 132 BLAGN are \tfm-detected with $S_{24 \, \rm \mu m}> 0.30 \, \rm mJy$. The histograms in the left panel of figure \ref{fig_agn} show the redshift distributions of the two samples, both of which include objects up to $z>3$.

The comparison of both redshift distributions shows that, at least up to $z\approx2$, the fraction of  $I<22.5$ BLAGN that are IR-detected has little variation with redshift. This fact indicates that the sensitivity limit  of the \tfm survey is not the only factor that determines the IR detection of these sources. Other physical differences might mean that some BLAGN are not IR-detected, even when lying at lower redshifts than others that are detected.

Within the unification scheme for AGN (Antonucci~1993), broad lines are present in the spectra when the source is oriented face-on and the central gas clouds are visible. The surrounding dusty torus is responsible for the IR emission.  The flux at IR wavelengths depends on the incoming flux of UV photons from the central engine, so it is related to the amount of material accreted onto the central black hole. Thus, the difference in the IR emission of these type-1 AGN could be directly indicating the accreting power in these objects.

As expected,  the most luminous IR sources in our flux-limited \tfm sample are AGN at redshifts $z>1$ (cf. figure \ref{fig_l24}), particularly BLAGN.  AGN are known to dominate the bright-end of the mid-IR luminosity function at high redshifts (Caputi et al.~2007). To investigate the presence of  other obvious (narrow-line) AGN within our sample, we looked for X-ray counterparts with $L_{X}>10^{42.5} \, \rm erg \, s^{-1}$ in the {\em XMM-Newton} maps of the COSMOS field (Hasinger et al.~2007). We found 35 IR sources associated with narrow-line $L_{X}>10^{42.5} \, \rm erg \, s^{-1}$ sources. The distribution of \tfm luminosities \nLntfm of these galaxies compared to the distribution of luminosities for the BLAGN is shown in the right panel of figure \ref{fig_agn}. It is clear from these two distributions that the BLAGN in our sample are considerably more luminous in the IR than the most powerful narrow-line AGN.

The fact that all of the identified most-luminous IR AGN are BLAGN might be partially  due to the optical magnitude cut limiting the zCOSMOS-bright survey. Some narrow-line AGN at high redshifts could also be IR-luminous, but they are probably too faint in the optical bands to be zCOSMOS targets. The prevalence of BLAGN among the most-luminous IR galaxies detected in zCOSMOS-bright could also be consequence of the orientation effect: in BLAGN, the optical nucleus is more exposed than in edge-on-oriented active sources.

On the other hand, 45 of our \tfm sources are characterised by an IRAC power-law SED, as defined in Section \ref{sec_rflum}. 30 out of 45 of these sources are also zCOSMOS-bright BLAGN. 
To inspect whether the remaining 15 IRAC power-law sources had also an independent signature of AGN activity, we looked for counterparts of these sources in the X-ray catalogues for the COSMOS field (Hasinger et al.~2007). We found that 11 out of 15 of the IRAC power-law sources not identified as zCOSMOS BLAGN are detected in X-rays. 
 
\begin{figure}
\epsscale{0.80}
\plotone{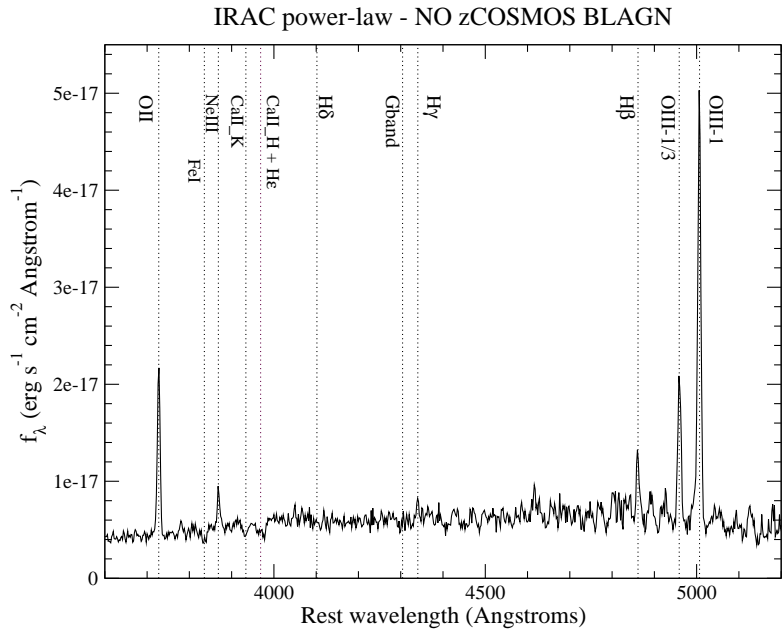}
\caption[]{\label{fig_irpl} Composite spectrum of 8 IRAC power-law SED \tfm sources at $0.5<z<0.85$ whose individual spectra are not classified as BLAGN.}
\end{figure}

 The composite zCOSMOS spectra of these objects also suggests the presence of active sources. Figure \ref{fig_irpl} shows the composite average spectrum of 8 IRAC power-law sources at $0.5<z<0.85$ for which the individual zCOSMOS spectra do not correspond to BLAGN. We see in this spectrum  high \oiii/\hb and \oiii/\oii line ratios and also the presence of [NeIII] in emission. All these characteristics suggest that most of the IRAC power-law sources which are not individually identified with zCOSMOS BLAGN are either narrow-line AGN or composite AGN/starburst systems.

\section{The infrared phase within the galaxy star-formation history}
\label{sec_irph}

Optical spectra have also features that provide information on the past star-formation history (SFH) of galaxies. For example, the presence of  higher-order Balmer lines in absorption (see figures \ref{fig_stack1} to \ref{fig_stack3}) indicates recent activity, i.e. star formation occurred within the last Gyr (e.g. Dressler \& Gunn 1983). Recently, Kauffmann et al.~(2003b) studied $4000 \, \rm \AA$-break strengths $\rm D_n(4000)$ and $\rm H\delta$ EW for a sample of local SDSS galaxies and showed how these two quantities can be used to put constraints on the galaxy mean stellar age and SFH. 

\begin{figure}
\epsscale{0.65}
\plotone{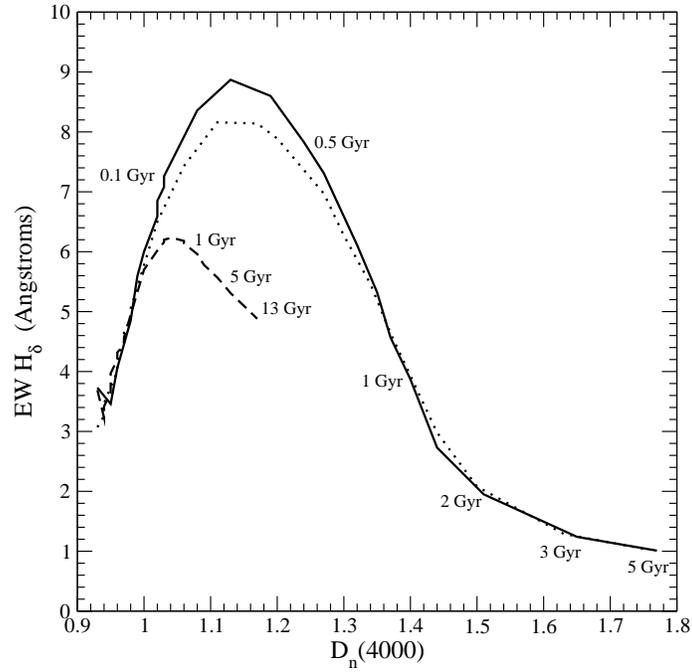}
\caption[]{\label{fig_modon}  The expected evolution of $\rm H\delta$ versus $\rm D_n(4000)$ for galaxies with different SFH: exponentially-declining  (with $\tau=0.01$ and $0.1 \, \rm Gyr$; solid and dotted lines, respectively) and a constant star formation (dashed line). The evolution has been measured using the Bruzual \& Charlot (Bruzual~2007) synthetic templates.}
\end{figure}

Figure \ref{fig_modon} shows the expected evolution of the $\rm H\delta$ EW versus $\rm D_n(4000)$ as a function of age for galaxies with different SFHs (in this section, positive values of the EW indicate line absorption). We modelled this evolution using the Bruzual \& Charlot~(Bruzual~2007)  templates for galaxies with exponentially-declining  star formations (with $\tau=0.01$ and $0.1 \, \rm Gyr$; solid and dotted lines, respectively) and a constant star formation (dashed line). We measured $\rm D_n(4000)$ as the ratio of the average flux densities between the $(4000-4100)$ and $(3850-3950) \, \rm \AA$ bands (Balogh et al.~1999).
We only considered models with solar metallicities, as both $\rm D_n(4000)$ and the $\rm H\delta$ EW have little dependence on metallicity at ages $<1 \, \rm Gyr$ after the onset of star formation (see Kauffmann et al.~2003b). As we show below, these are the timescales of relevance for IR galaxies.

Our aim is then to locate IR galaxies in the $\rm H\delta$ EW - $\rm D_n(4000)$ diagram and see whether we can constrain at what stage of a galaxy life the IR phase is occurring.

\begin{figure}
\epsscale{0.65}
\plotone{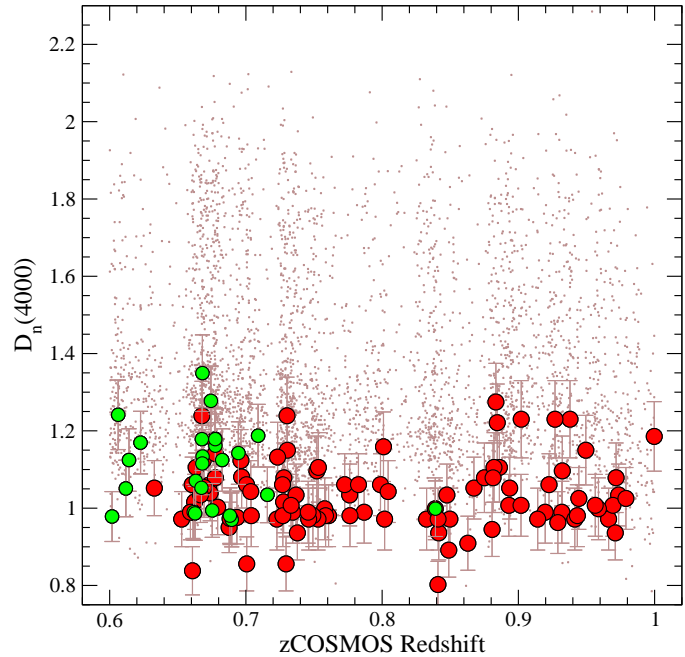}
\caption[]{\label{fig_d4000}  $\rm D_n(4000)$ values for \tfm galaxies as a function of redshift. Brown dots indicate all the other zCOSMOS galaxies in the same redshift range (Lamareille et al., in preparation).}
\end{figure}

The measurement of $\rm D_n(4000)$ can be easily done on individual spectra. We show in figure \ref{fig_d4000} the resulting $\rm D_n(4000)$ measurements for all our \tfm galaxies at $0.6<z<1.0$. The zCOSMOS BLAGN have been excluded from this analysis.  We corrected the $\rm D_n(4000)$ measurements for dust extinction, although these corrections are very small, given the narrow wavelength range considered. For a comparison, we show the $\rm D_n(4000)$ measurements for all the non-IR-detected zCOSMOS galaxies in the same redshift range  (brown dots in figure \ref{fig_d4000}; Lamareille et al., in preparation).

We see from figure \ref{fig_d4000} that the $\rm D_n(4000)$ strengths of IR-selected galaxies are in all cases quite small ($\rm D_n(4000) \lsim 1.3$), and they do not depend on redshift. The values we obtain for $\rm D_n(4000)$ in \tfm galaxies are in agreement with those measured by Marcillac et al.~(2006) on a sample of 25 {\em ISO}-selected LIRGs at $z\sim0.7$. The small values for $\rm D_n(4000)$ indicate the presence of  young generations of stars in IR galaxies. We note that IR galaxies display the smallest $\rm D_n(4000)$ values among zCOSMOS galaxies, for which $\rm D_n(4000)$ can take values up to 2 or more.

In fact, the existence of young stars is expected in galaxies that are IR-selected, i.e. that are selected to have on-going star formation.  However, an old ($> 1 \, \rm Gyr$) galaxy with on-going declining star formation displays quite redder colours. The fact that $\rm D_n(4000)$ is invariably small indicates that IR galaxies either are young systems or are experiencing a rejuvenation of their stellar content.

Unfortunately, the  $\rm H\delta$ EW are quite small and the associated errors too large  as to rely on individual measurements in most cases.  We then measured $\rm H\delta$ only on average stacked spectra, for which the signal-to-noise ratio is much higher. We constructed 11 composites corresponding to  galaxies of similar IR luminosities in narrow redshift bins between $z=0.6$ and 1.0. Each composite is the average of 17 to 32 individual spectra, depending on the case.

\begin{figure}
\epsscale{0.75}
\plotone{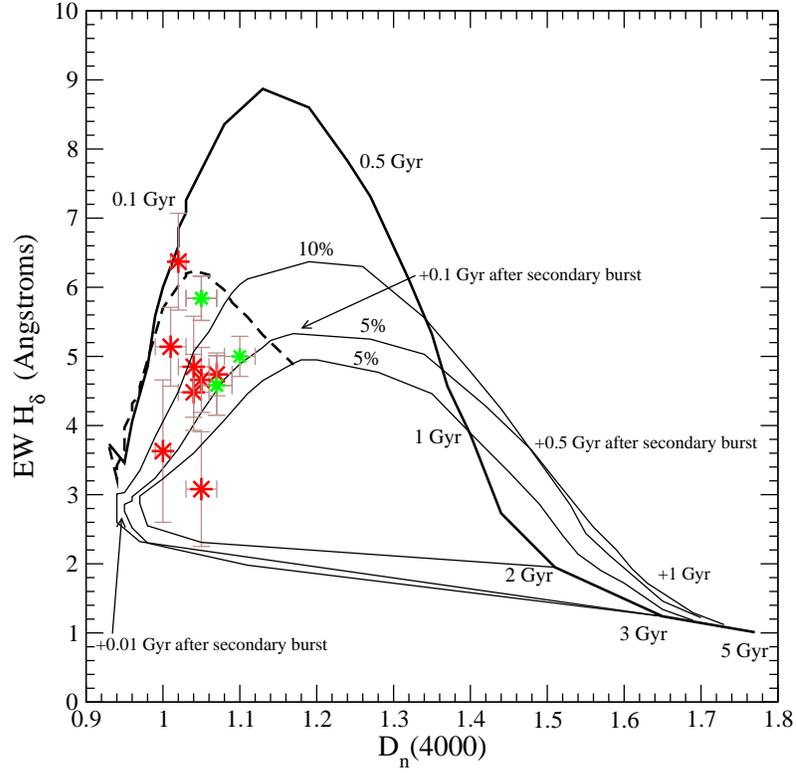}
\caption[]{\label{fig_secb} The average location of \tfm galaxies in the $\rm H\delta$ versus $\rm D_n(4000)$ diagram. The colour and size code for the asterisks is the same as in previous figures. Thick lines represent the same modelled evolution shown in figure \ref{fig_modon}. The thin solid lines show the effect of secondary bursts of star formation that produce 5 or 10\%  of the already-assembled stellar mass. We show examples where the burst starts when the galaxy is 2 or 3 Gyrs old. The numbers indicated with a plus sign in front correspond to the time passed since the onset of the secondary burst.}
\end{figure}

To correct the $\rm H\delta$ EW  for line filling, we  measured the $\rm H\gamma$ emission line  on each composite, corrected it for stellar absorption and assumed an intrinsic decrement $\rm H\gamma/H\delta=1.80$ (for a case B recombination with temperature $T=10,000 \rm K$; Osterbrock~1989). Beyond $z\sim0.9$,  $\rm H\gamma$ is within the fringing-affected region of our spectra and its measurement becomes quite uncertain. Thus, at these high redshifts, we simply used the same line-filling corrections as in immediately lower redshift bins. The resulting  average $\rm H\delta$ EW versus $\rm D_n(4000)$ measurements are shown with asterisks in figure \ref{fig_secb}.

IR galaxies occupy, on average, a well-defined place in the  $\rm H\delta$ EW - $\rm D_n(4000)$ diagram. The ideal single constant SFH is consistent with the average location of some IR galaxies but, still in this  case, the associated ages are very young ($< 1 \, \rm Gyr$). For most IR galaxies, however, the track of a constant SFH is well above their location in the $\rm H\delta$ EW - $\rm D_n(4000)$ diagram.

 The locus of IR galaxies in the $\rm H\delta$ EW - $\rm D_n(4000)$ plane can, instead, be well explained by secondary bursts of star formation (thin solid lines in figure \ref{fig_secb}; see also Hammer et al.~2005; Marcillac et al.~2006).  These secondary bursts only need to form an additional minor amount (5-10\%) of stellar mass in these galaxies to produce the $\rm H\delta$ EW and $\rm D_n(4000)$ values observed. Here we confirm that such a scenario can explain the average properties of  the brightest IR galaxies found all the way  from redshifts $z=0.6$ through 1.0.  

Two additional interesting conclusions can be extracted from  figure \ref{fig_secb}:  firstly, the small $\rm D_n(4000)$ values and relatively small  $\rm H\delta$ EW displayed by some composites indicates that the rejuvenation of the stellar populations occurs, in many cases, in relatively old ($> 1 \, \rm Gyr$) galaxies.  We know that, at $z<1$, LIRGs and ULIRGs are generally associated with intermediate-to-large stellar mass galaxies ($M \gsim 1 \times 10^{10} \, \rm M_{\odot}$; see Caputi et al.~2006a). So, the brightest IR emission at these redshifts is being hosted by galaxies which already formed most of their stellar mass at earlier epochs.

Secondly, the average location of our sources on the $\rm H\delta$ EW - $\rm D_n(4000)$ plane strongly constrains when the IR phase is being produced:  on average, galaxies are IR bright in the elapsed time between $10^7$ and $10^8$ years after the onset of the secondary burst. We note that our results do not completely exclude the possibility that some galaxies have more prolonged IR phases characterised by more or less constant SFR. Galaxies that still have sufficiently large gas reservoirs remaining at $z<1$ could in principle sustain the typical SFR of LIRGs  ($\sim 10-100 \, \rm M_{\odot}/yr$) for some longer times. However, given the already-assembled stellar  masses of LIRGs and ULIRGs, this situation is probably implausible in most cases.

Most of the dust contained in IR galaxies is produced by AGB stars, which start to inject dust into the interstellar medium only a few hundreds of Myr after the production of the burst (e.g. Dwek~2005). The fact that the IR phase occurs between $10^7$ and $10^8$ years after the onset of the secondary burst indicates that the dust being heated during the IR phase is probably `old dust', i.e. dust that has been created by previous generations of stars.

\section{Young powerful starbursts or further AGN activity?}
\label{sec_young}

 We have seen in the previous section that the bulk of the IR phase in galaxies at $z<1$ is observed, on average, $10^7 - 10^8$ years after the onset of a secondary burst of star formation.  The minimum  average time for the detection of the IR phase is in part simply due to the lower rate of galaxies that can be detected in shorter ($<10^7$ yr) elapsed times. But, if this were the only factor determining the rate of detection of younger starbursts, one would possibly expect to find around one out of ten IR galaxies being at the earliest stages of a new burst of star formation.
  
  We looked for  galaxies with signatures of young ($<10^7$ yr) starbursts in their optical spectra within our \tfm sample.   At  early stages of star formation, the EW of  absorption features associated with high-order Balmer lines is expected to be small (cf. figure \ref{fig_secb}). In addition, star formation should be sufficiently strong as to clearly detect even the higher-order Balmer lines in emission (except, maybe, for extremely-obscured sources).   As we explained in Section \ref{sec_irph}, the typical EW widths measured for $\rm H\delta$ on different spectra are, in general, small and the associated errors quite large as to consider these measurements on an individual basis. In spite of this, we used these tentative measurements of $\rm H\delta$ to look for  candidate galaxies to be hosts of very young starbursts.

\begin{figure}
\plotone{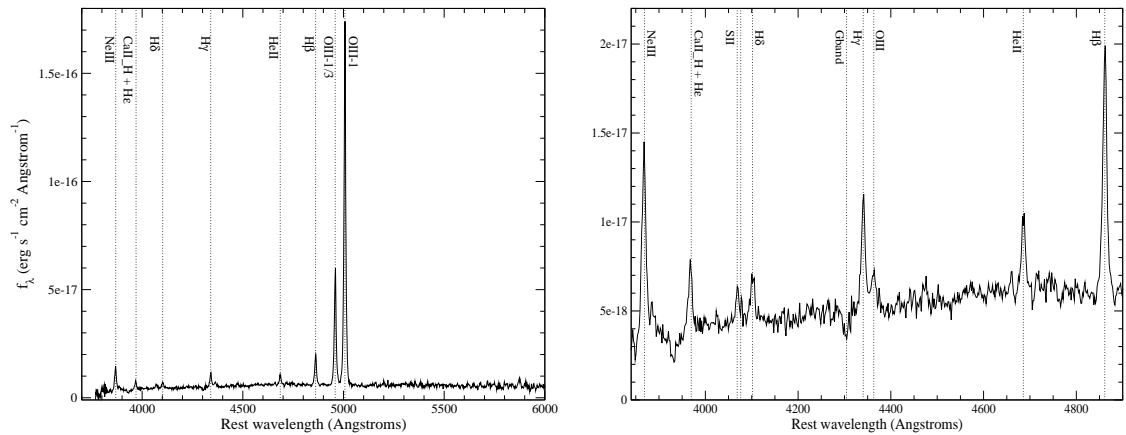}
\caption[]{\label{fig_young} Example of a \tfm $\,\,$ galaxy candidate to be at the earliest stages of a burst of star formation. Left panel: spectrum over the total observed wavelength range. Right panel: detailed view of the region of high-order Balmer lines.}
\end{figure}

We found four likely candidates for young starbursts within our \tfm sample. The spectra of all of these galaxies are characterised by the presence of $\rm H\delta$ and $\rm H\varepsilon$ in emission. One case also shows a slightly broad HeII line (see example in figure \ref{fig_young}), which is a signature of stellar winds originated in Wolf-Rayet stars in young-starburst galaxies (Schaerer et al.~1999).

We  note, however, that other characteristics of these spectra could also suggest the possibility of AGN activity: in particular, the presence of a [NeIII] emission line or the high \oiii/\oii ratios. Actually, both star formation and AGN activity are likely to coexist in these objects. Unfortunately, the lack of spectral coverage for the \ha and \nii lines prevents us from further constraining the nature of these sources.

We looked for X-ray counterparts of these objects in the {\em XMM-Newton} catalogue for the COSMOS field (Hasinger et al.~2007). None of them appear to be identified with an X-ray source. These galaxies are not identified either in the deeper {\em Chandra Observatory} X-ray maps for the COSMOS field (C-COSMOS; P.I. M. Elvis).

We found three more galaxies in our sample that also have $\rm H\delta$ and $\rm H\varepsilon$ in emission, but the lines are weaker and the noise in the spectra is higher. Some extra candidates are probably missing because of extinction effects. Still, these presumably young starbursts seem to be quite less than $\sim$10\% of our galaxy sample. This fact suggests that the onset of the IR phase within  $\sim 10^7$ yr after the burst might be indeed a rare phenomenon. Of course, given the optical magnitude cut of the zCOSMOS survey, we cannot exclude that IR galaxies at the very earliest stages of a rejuvenating starburst are more common among optically faint sources.

The low fraction of very young starbursts among IR-selected galaxies is a known phenomenon in the local Universe.  Roussel et al.~(2006) studied the case of  NGC 1377, an extremely young ($\lsim 1 \, \rm Myr$) local starburst, selected for being an outlier in the far-IR/radio correlation. These authors note that such galaxies represent $\sim1\%$ of faint {\em IRAS} galaxies. In contrast to our cases, the optical spectrum of NGC 1377 shows very weak  \oii, \oiii and \hb emission, due to dust obscuration. The mid-IR spectrum of this source, instead, is dominated by  molecular-hydrogen $H_2$ lines, indicating a young reservoir of star formation. 

In a forthcoming paper, we will analyse the presence of young starburst candidates within the deeper \tfm galaxy sample for the COSMOS field up to higher redshifts.  This will allow us to constrain whether a fainter IR phase is more often associated with these sources at the earliest stages of star formation.

\section{Summary and Conclusions}
\label{sec_concl}

In this work, we have presented the results of the study of zCOSMOS optical spectra  for 609 \tfm galaxies with $S_{24 \, \rm \mu m}> 0.30 \, \rm mJy$, selected over 1.5 deg$^2$ of the COSMOS field.  This is the most extensive analysis of the spectra of a very large  sample of mid-IR galaxies analysed to date that covers the redshift ranges $0<z \lsim 1$ and $0<z \lsim 3$ for normal galaxies and AGN, respectively.

Depending on the lines present in the spectra, we made different diagnostics for sources at different redshifts. As expected, at low $z<0.3$ redshifts, our sample is dominated by IR-normal galaxies (in our classification, strictly, \nLnn). The optical spectra of these sources display a large variety of emission-line EW and ratios. This is in part due to a spatial sampling  effect, because the 1-arcsec-width VIMOS slit maps only the central region within each galaxy, which in some cases is not representative of the entire galaxy. The same effect has been observed in the study of low-redshift {\em IRAS} galaxy optical spectra.  But the variety of spectral characteristics also reveals the different physical conditions  of the hosts of normal IR galaxies. This class includes galaxies with different degrees of chemical evolution and already-assembled stellar masses:  from some low-mass, low-metallicity galaxies to many already massive, chemically enriched,  galaxies.

At higher redshifts,  we studied the variation of the ionisation levels in different luminosity IR galaxies, through their  \oiii/\hb line ratios.  We found that the average \oiii/\hb ratio for  \nLnu galaxies at $z\sim 0.6-0.7$ is more than 0.6 dex higher than the average ratio of all zCOSMOS galaxies at similar redshifts. This allows us to conclude that ULIRGs have distinct optical spectral properties when compared to typical optically-selected sources.  In particular, for some of our galaxies, we find very high $\log_{10}($\oiii$/$\hb$) > 0.5$ ratios, comparable to those of sub-millimetre galaxies at $z>1.5$. These are cases where massive star formation and narrow-line AGN activity are likely to coexist. Near or mid-IR  spectroscopy is necessary to provide further information on the ionisation source in these galaxies.

Around $1/3$ of  the galaxies for which we could derive  metallicities at $0.5<z<0.7$ are unambiguously  below the general mass-metallicity relation at the corresponding redshift. Although, in some cases, the plausible presence of an AGN could affect the derived metallicities, this is probably not the main cause of the  discrepancies with the mass-metallicity relation. Instead, gas mixing due to inflows and outflows during the IR phase could temporarily move sources out of the mass-metallicity trend (see e.g. Rupke et al.~2008). 

The comparison of optical emission-line derived and fiducial (IR+UV) SFR indicates that IR galaxies are generally quite affected by extinction. SFR derived from the extincted \ha and \hb lines  underestimate the real SFR, on average, by  factors $\sim5$ and $\sim10$, respectively.  We note, however, that these discrepancies are expected to be maximal for the most luminous IR galaxies. In other galaxies, the \ha or \hb luminosities should constitute more reliable SFR estimators. 

Also from our spectroscopic analysis, we found that it is difficult to explain the properties of many IR galaxies at $z<1$ with a constant SFH. Instead, a SFH characterised by secondary bursts of star formation can simultaneously explain the blue continuum and relatively small $H\delta$ EW observed in most of these sources (see also Hammer et al.~2005).  These results also constrain the epoch of IR activity: the LIRG and ULIRG phases for star-forming galaxies occurs, on average, between $10^7$ and $10^8$ years after the onset of the secondary burst. Most of the hosts of intense IR activity at $0.6<z<1.0$ have older underlying stellar populations and have assembled  most of their stellar mass in the past (Caputi et al.~2006a). The very short times for the onset of the IR phase after the burst also suggest that the dust being heated during these rejuvenating episodes of star formation must be `old dust', i.e. dust produced by AGB stars from previous stellar generations. All these conclusions are valid in general for all the galaxies we have analysed here, independently of their IR luminosity and redshift, suggesting that the way in which star formation proceeds is basically the same in most of these systems (i.e. independently of the mechanism triggering the burst).

Finally, $\sim$10\% of our galaxies are recognised as BLAGN from their zCOSMOS spectra. Another $\sim 5-6\%$ are X-ray-luminous AGN with narrow-line spectra.  These figures are consistent with the fraction of AGN-dominated IR sources found in the literature (e.g. Alonso-Herrero et al.~2006; Caputi et al.2006a,b,2007). However, AGN activity might also be present in some other galaxies, as suggested by their high ionisation levels or {\em IRAC}-power law SEDs. The exact fraction of AGN and AGN/star-forming composite systems among IR galaxies is still far from being clear. In the future, very deep X-ray data and far-IR data obtained with the {\em Herschel} telescope will be available  for the COSMOS field. Both datasets analysed in conjunction should help to put stronger constraints on the AGN contribution to the IR background.

 The results obtained in this paper offer some important clues of the role of the IR phase in star-formation history and galaxy evolution, mainly at $z<1$. IR sources are the signposts of the most intense star formation activity and on-going build-up of stellar mass. But the IR phase is a kind of a transitory state in which  some galaxy properties (e.g. the strength of the $\rm 4000 \, \AA$ break and, in some cases, the metallicity) are perturbed from their regular values in the  IR-inactive galaxy life. This should be kept in mind when considering IR sources for a census of galaxy properties, as the values measured for some of these properties could be well apart from those that will hold at long term once the IR phase has faded out.

\acknowledgments

This paper is based on observations made with the VIMOS spectrograph on the Melipal-VLT telescope, undertaken at the European Southern Observatory (ESO) under Large Program 175.A-0839. Also based on observations made with the {\em Spitzer} Observatory, which is operated by the Jet Propulsion Laboratory, California Institute of Technology, under NASA contract 1407. 

We thank our referee Jonathan Gardner for useful comments and suggestions. We are grateful to Gustavo Bruzual and St\'ephane Charlot for making available the latest version of their synthetic SED template library.

%

\clearpage

%
\begin{deluxetable}{lr} 
\tablewidth{15cm} 
\tablecaption{The numbers of \tfm galaxies with $S_{24 \, \rm \mu m}> 0.30 \, \rm mJy$ which satisfy the different selection criteria imposed by the zCOSMOS-bright survey. Each category is included in the previous one. All associations have been done within a 2 arcsec matching distance.
\label{tab_galsel}} 
\tablehead{  
\colhead{Selection criterion} &
\colhead{Number of sources} 
} 
\startdata 
Sources in SCOSMOS shallow ($1.75\times1.97$ deg$^2$) & 9,807 \\
Sources in the zCOSMOS-bright 10k-sample field (1.5 deg$^2$) & 3,150 \\
Sources with $I<22.5$ AB mag & 2,084\\
Sources with associations in zCOSMOS-bright 10k sample & 703 \\
Secure or likely zCOSMOS-bright associations & 668 \\
Sources with secure zCOSMOS-bright spectra & 611 \\
\enddata 
\end{deluxetable}

%
\begin{deluxetable}{lrrrc} 
\tablewidth{12cm} 
\tablecaption{The \hb and \oiii line fluxes  for the galaxies in our sample for which both line EW are greater than $5 \, \rm \AA$.
All fluxes are aperture and extinction-corrected. \hb fluxes also include corrections for stellar absorption. The flux units are $10^{-15} \, \rm erg \, s^{-1} cm^{-2}$. zCOSMOS BLAGN have been excluded. The sources flagged with an asterisk correspond to X-ray AGN with $L_{X}>10^{42.5} \, \rm erg \, s^{-1}$.
\label{tab_hboiii}} 
\tablehead{  
\colhead{zCOSMOS id} &
\colhead{$z_{spec}$} &
\colhead{\hb} &
\colhead{\oiii}&
\colhead{$\log_{10}$\nLntfm}
} 
\startdata 
813537	&	0.2602		&	1.74$\pm$0.42	&	1.34$\pm$0.41	&	$<10$	\\
812917	&	0.3398		&	1.41$\pm$0.28	&	1.35$\pm$0.25	&	$<10$	\\
812047	&	0.2648		&	3.26$\pm$0.57	&	16.31$\pm$2.30	&	$<10$	\\
818329	&	0.2656		&	3.16$\pm$0.56	&	12.90$\pm$1.83	&	$<10$	\\
825619	&	0.3100		&	2.30$\pm$0.49	&	1.72$\pm$0.38	&	$<10$	\\
817135	&	0.3220		&	1.79$\pm$0.39	&	4.48$\pm$0.71	&	$<10$	\\
822887	&	0.2167		&	4.56$\pm$0.92	&	5.66$\pm$0.89	&	$<10$	\\
840778	&	0.2614		&	2.79$\pm$0.70	&	1.84$\pm$0.55	&	$<10$	\\
848386	&	0.2861		&	2.29$\pm$0.56	&	0.90$\pm$0.22	&	$<10$	\\
831576	&	0.2512		&	1.96$\pm$0.42	&	0.94$\pm$0.21	&	$<10$	\\
841054	&	0.2726		&	1.91$\pm$0.48	&	1.38$\pm$0.44	&	$<10$	\\
840771	&	0.3302		&	1.65$\pm$0.36	&	0.83$\pm$0.17	&	$<10$	\\
836248	&	0.2602		&	2.66$\pm$0.57	&	1.75$\pm$0.32	&	$<10$	\\
837327	&	0.2190		&	5.33$\pm$1.06	&	4.03$\pm$0.66	&	$<10$	\\
826959	&	0.2607		&	1.73$\pm$0.39	&	2.46$\pm$1.00	&	$<10$	\\
844388	&	0.2158		&	1.99$\pm$0.41	&	4.02$\pm$0.61	&	$<10$	\\
814007	&	0.5866		&	1.75$\pm$0.75	&	1.34$\pm$0.28	&	10-11	\\
801253	&	0.4803		&	2.88$\pm$0.55	&	2.28$\pm$0.36	&	10-11	\\
812879	&	0.2506		&	7.73$\pm$1.35	&	26.64$\pm$4.05	&	10-11	\\
814176	&	0.3753		&	1.96$\pm$0.42	&	1.38$\pm$0.28	&	10-11	\\
813806	&	0.6896		&	1.65$\pm$0.67	&	2.79$\pm$1.73	&	10-11	\\
820021	&	0.6753		&	1.81$\pm$0.50	&	0.52$\pm$0.27	&	10-11	\\
826685	&	0.6018		&	2.42$\pm$0.56	&	1.75$\pm$0.33	&	10-11	\\
825318	&	0.4789		&	0.98$\pm$0.21	&	1.28$\pm$0.28	&	10-11	\\
817189	&	0.4609		&	3.10$\pm$0.69	&	36.97$\pm$5.32	&	10-11	\\
823751	&	0.4795		&	2.73$\pm$0.58	&	2.54$\pm$0.43	&	10-11	\\
827923	&	0.4332		&	1.37$\pm$0.30	&	1.38$\pm$0.25	&	10-11	\\
834174	&	0.5027		&	1.69$\pm$0.47	&	2.33$\pm$0.51	&	10-11	\\
833601	&	0.3342		&	2.49$\pm$0.55	&	1.54$\pm$0.28	&	10-11	\\
824223	&	0.6776		&	1.83$\pm$0.63	&	1.58$\pm$0.69	&	10-11	\\
832126	&	0.3711		&	3.18$\pm$0.62	&	5.32$\pm$0.81	&	10-11	\\
831349	&	0.3115		&	4.43$\pm$0.89	&	3.56$\pm$0.56	&	10-11	\\
835952	&	0.3616		&	2.31$\pm$0.52	&	1.16$\pm$0.25	&	10-11	\\
844837	&	0.5021		&	1.92$\pm$0.41	&	1.97$\pm$0.32	&	10-11	\\
843329	&	0.5064		&	2.56$\pm$0.46	&	13.91$\pm$2.02	&	10-11	\\
804831	&	0.3466		&	2.69$\pm$0.56	&	1.32$\pm$0.24	&	10-11	\\
834565	&	0.5339		&	1.82$\pm$0.54	&	2.23$\pm$0.36	&	10-11	\\
826188	&	0.3453		&	3.93$\pm$0.75	&	3.71$\pm$0.65	&	10-11	\\
816998$^\ast$ &	0.4248		&	4.86$\pm$1.03	&	8.62$\pm$1.36	&	10-11	\\
838297	&	0.5037		&	1.22$\pm$0.36	&	0.91$\pm$0.21	&	10-11	\\
835862	&	0.4027		&	3.32$\pm$0.69	&	1.78$\pm$0.34	&	10-11	\\
844486	&	0.4694		&	1.93$\pm$0.40	&	0.83$\pm$0.21	&	10-11	\\
851898	&	0.6655		&	2.02$\pm$0.51	&	4.57$\pm$0.83	&	10-11	\\
826453	&	0.3547		&	2.76$\pm$0.56	&	1.12$\pm$0.28	&	10-11	\\
803488	&	0.5867		&	4.69$\pm$1.79	&	39.11$\pm$9.32	&	$>11$	\\
820949	&	0.5712		&	3.00$\pm$0.60	&	2.23$\pm$0.45	&	$>11$	\\
812432	&	0.6611		&	7.59$\pm$2.38	&	39.35$\pm$11.11	&	$>11$	\\
817871	&	0.6738		&	2.42$\pm$0.53	&	14.74$\pm$3.06	&	$>11$	\\
841690	&	0.4821		&	11.01$\pm$1.92	&	58.98$\pm$8.40	&	$>11$	\\
831770	&	0.6877		&	3.35$\pm$0.61	&	3.44$\pm$0.63	&	$>11$	\\
830349	&	0.6693		&	1.94$\pm$0.48	&	1.59$\pm$0.42	&	$>11$	\\
838689$^\ast$ &	0.6884		&	10.84$\pm$6.62	&	24.96$\pm$5.88	&	$>11$	\\
836868	&	0.6793		&	2.42$\pm$0.61	&	23.14$\pm$3.16	&	$>11$	\\
851888	&	0.6667		&	2.27$\pm$0.88	&	3.07$\pm$0.97	&	$>11$	\\
820787	&	0.6753		&	3.25$\pm$0.58	&	26.00$\pm$3.72	&	$>11$	\\
\enddata 
\end{deluxetable} 

\clearpage

%
\begin{longtable}{lrrrc} 
\tablewidth{12cm} 
\tablecaption{The \ha and \nii line fluxes  for the galaxies in our sample for which both line EW are greater than $5 \, \rm \AA$.
All fluxes are aperture and extinction-corrected. \ha fluxes also include corrections for stellar absorption. The flux units are $10^{-15} \, \rm erg \, s^{-1} cm^{-2}$. zCOSMOS BLAGN have been excluded. The source flagged with an asterisk corresponds to an X-ray AGN with $L_{X}>10^{42.5} \, \rm erg \, s^{-1}$.
\label{tab_hanii}} 
\tablehead{  
\colhead{zCOSMOS id} &
\colhead{$z_{spec}$} &
\colhead{\ha} &
\colhead{\nii}&
\colhead{$\log_{10}$\nLntfm}
} 
\startdata 
806808	&	0.1267		&	7.22$\pm$1.30	&	5.90$\pm$1.19	&	$<10$	\\
805853	&	0.2823		&	4.37$\pm$0.59	&	2.14$\pm$0.55	&	$<10$	\\
813332	&	0.1662		&	11.64$\pm$1.33	&	3.97$\pm$0.62	&	$<10$	\\
812999	&	0.1330		&	15.52$\pm$2.08	&	7.23$\pm$1.50	&	$<10$	\\
805921	&	0.2673		&	9.03$\pm$1.09	&	2.98$\pm$0.91	&	$<10$	\\
805095	&	0.1848		&	11.21$\pm$1.30	&	2.69$\pm$0.47	&	$<10$	\\
800509	&	0.2061		&	9.92$\pm$1.48	&	4.21$\pm$1.00	&	$<10$	\\
812989	&	0.1335		&	19.98$\pm$2.34	&	8.77$\pm$1.41	&	$<10$	\\
812793	&	0.2202		&	5.44$\pm$1.12	&	4.33$\pm$1.51	&	$<10$	\\
812559	&	0.1868		&	6.00$\pm$0.77	&	2.98$\pm$0.52	&	$<10$	\\
820251	&	0.0922		&	55.28$\pm$6.46	&	26.53$\pm$4.12	&	$<10$	\\
819521	&	0.1089		&	16.21$\pm$2.03	&	9.96$\pm$1.68	&	$<10$	\\
811456	&	0.2835		&	8.02$\pm$2.59	&	6.10$\pm$1.79	&	$<10$	\\
819880	&	0.0928		&	24.92$\pm$3.35	&	12.57$\pm$2.23	&	$<10$	\\
819278	&	0.1692		&	5.10$\pm$0.61	&	2.03$\pm$0.34	&	$<10$	\\
812047	&	0.2648		&	9.40$\pm$1.04	&	0.18$\pm$0.04	&	$<10$	\\
819178	&	0.1240		&	6.68$\pm$1.03	&	3.34$\pm$0.75	&	$<10$	\\
811299	&	0.1689		&	25.69$\pm$3.21	&	12.79$\pm$2.07	&	$<10$	\\
826189	&	0.2203		&	7.04$\pm$0.85	&	1.60$\pm$0.34	&	$<10$	\\
811017	&	0.0994		&	6.98$\pm$0.86	&	1.88$\pm$0.65	&	$<10$	\\
818329	&	0.2656		&	9.10$\pm$1.18	&	0.82$\pm$0.28	&	$<10$	\\
826438	&	0.2826		&	5.92$\pm$3.42	&	2.61$\pm$0.97	&	$<10$	\\
826635$^\ast$ &	0.1333		&	21.68$\pm$2.85	&	12.21$\pm$2.27	&	$<10$	\\
818715	&	0.1230		&	16.72$\pm$2.18	&	7.16$\pm$1.43	&	$<10$	\\
818641	&	0.1945		&	17.46$\pm$2.03	&	7.32$\pm$1.19	&	$<10$	\\
825585	&	0.2814		&	5.58$\pm$1.09	&	4.22$\pm$1.67	&	$<10$	\\
825178	&	0.2504		&	8.38$\pm$1.65	&	5.28$\pm$1.12	&	$<10$	\\
818115	&	0.2200		&	9.06$\pm$1.66	&	4.53$\pm$1.38	&	$<10$	\\
824542	&	0.1056		&	12.73$\pm$1.50	&	3.77$\pm$0.67	&	$<10$	\\
825011	&	0.1873		&	19.71$\pm$2.35	&	11.50$\pm$2.06	&	$<10$	\\
823948	&	0.1694		&	8.43$\pm$0.99	&	3.56$\pm$0.62	&	$<10$	\\
824741	&	0.2497		&	5.38$\pm$0.64	&	1.37$\pm$0.22	&	$<10$	\\
824137	&	0.1101		&	34.82$\pm$4.55	&	27.58$\pm$4.58	&	$<10$	\\
824379	&	0.0931		&	8.06$\pm$0.94	&	2.65$\pm$0.43	&	$<10$	\\
824401	&	0.2143		&	8.97$\pm$1.13	&	4.22$\pm$0.70	&	$<10$	\\
823372	&	0.1848		&	8.86$\pm$1.23	&	4.24$\pm$0.77	&	$<10$	\\
823895	&	0.2478		&	8.69$\pm$1.51	&	3.76$\pm$0.94	&	$<10$	\\
822887	&	0.2167		&	13.16$\pm$1.49	&	4.38$\pm$0.89	&	$<10$	\\
826778	&	0.1861		&	5.76$\pm$0.71	&	2.26$\pm$0.43	&	$<10$	\\
833704	&	0.1303		&	11.47$\pm$1.63	&	7.07$\pm$1.40	&	$<10$	\\
834384	&	0.1673		&	13.39$\pm$1.82	&	7.49$\pm$1.49	&	$<10$	\\
834432	&	0.1879		&	5.86$\pm$0.85	&	3.21$\pm$0.69	&	$<10$	\\
832637	&	0.2519		&	6.14$\pm$1.17	&	3.38$\pm$0.81	&	$<10$	\\
832356	&	0.0926		&	8.15$\pm$0.96	&	2.34$\pm$0.40	&	$<10$	\\
824674	&	0.1854		&	4.58$\pm$0.74	&	5.26$\pm$1.38	&	$<10$	\\
840778	&	0.2614		&	8.05$\pm$1.06	&	2.56$\pm$0.53	&	$<10$	\\
832229	&	0.1862		&	7.33$\pm$1.07	&	5.64$\pm$1.28	&	$<10$	\\
824103	&	0.1653		&	6.43$\pm$0.87	&	2.41$\pm$0.47	&	$<10$	\\
848386	&	0.2861		&	6.60$\pm$0.81	&	4.25$\pm$0.82	&	$<10$	\\
831755	&	0.2656		&	7.80$\pm$1.26	&	3.60$\pm$0.98	&	$<10$	\\
840553	&	0.1255		&	6.87$\pm$0.92	&	2.28$\pm$0.52	&	$<10$	\\
823563	&	0.0754		&	98.04$\pm$11.45	&	60.04$\pm$9.20	&	$<10$	\\
830561	&	0.1234		&	25.84$\pm$4.00	&	15.09$\pm$3.43	&	$<10$	\\
840320	&	0.2605		&	5.95$\pm$0.72	&	6.09$\pm$1.68	&	$<10$	\\
831576	&	0.2512		&	5.66$\pm$0.66	&	1.32$\pm$0.23	&	$<10$	\\
841054	&	0.2726		&	5.53$\pm$0.83	&	1.65$\pm$0.36	&	$<10$	\\
839733	&	0.2485		&	6.34$\pm$1.60	&	5.26$\pm$1.65	&	$<10$	\\
830791	&	0.2484		&	6.41$\pm$0.97	&	2.06$\pm$0.62	&	$<10$	\\
830588	&	0.1711		&	6.42$\pm$0.74	&	1.84$\pm$0.29	&	$<10$	\\
848030	&	0.1877		&	10.94$\pm$1.24	&	4.03$\pm$0.63	&	$<10$	\\
839005	&	0.2685		&	4.69$\pm$0.65	&	3.24$\pm$0.95	&	$<10$	\\
830461	&	0.2217		&	10.79$\pm$2.73	&	5.84$\pm$1.52	&	$<10$	\\
830823	&	0.1215		&	35.91$\pm$4.29	&	22.34$\pm$3.45	&	$<10$	\\
830514	&	0.1273		&	11.34$\pm$1.93	&	8.17$\pm$1.83	&	$<10$	\\
829482	&	0.1273		&	16.71$\pm$2.09	&	4.96$\pm$0.98	&	$<10$	\\
830643	&	0.2170		&	14.57$\pm$9.63	&	9.42$\pm$2.78	&	$<10$	\\
837726	&	0.2210		&	7.97$\pm$1.32	&	2.38$\pm$0.66	&	$<10$	\\
846444	&	0.2195		&	10.12$\pm$2.04	&	6.66$\pm$1.86	&	$<10$	\\
837325	&	0.2197		&	12.48$\pm$1.44	&	4.96$\pm$0.83	&	$<10$	\\
837332	&	0.1139		&	7.51$\pm$0.84	&	0.72$\pm$0.12	&	$<10$	\\
837461	&	0.2198		&	11.96$\pm$2.57	&	3.10$\pm$0.66	&	$<10$	\\
837233	&	0.2135		&	9.60$\pm$1.73	&	3.11$\pm$0.59	&	$<10$	\\
836248	&	0.2602		&	7.68$\pm$0.88	&	3.00$\pm$0.70	&	$<10$	\\
828432	&	0.1769		&	6.33$\pm$0.77	&	1.26$\pm$0.25	&	$<10$	\\
843243	&	0.1035		&	9.79$\pm$1.22	&	2.98$\pm$0.57	&	$<10$	\\
843517	&	0.0832		&	22.55$\pm$2.64	&	6.23$\pm$1.16	&	$<10$	\\
842703	&	0.1794		&	10.34$\pm$1.23	&	4.44$\pm$0.90	&	$<10$	\\
845465	&	0.1240		&	20.51$\pm$2.71	&	6.98$\pm$1.86	&	$<10$	\\
844830	&	0.0731		&	11.35$\pm$1.31	&	1.08$\pm$0.22	&	$<10$	\\
813014	&	0.1332		&	34.80$\pm$4.17	&	8.69$\pm$1.52	&	$<10$	\\
811073	&	0.1682		&	5.23$\pm$0.70	&	2.19$\pm$0.41	&	$<10$	\\
837327	&	0.2190		&	15.37$\pm$1.77	&	4.96$\pm$0.87	&	$<10$	\\
817508	&	0.1866		&	14.22$\pm$2.11	&	6.66$\pm$1.50	&	$<10$	\\
827229	&	0.2208		&	4.32$\pm$0.57	&	1.41$\pm$0.55	&	$<10$	\\
844388	&	0.2158		&	5.75$\pm$0.74	&	1.10$\pm$0.23	&	$<10$	\\
835421	&	0.1237		&	10.88$\pm$1.29	&	2.56$\pm$0.49	&	$<10$	\\
812879	&	0.2506		&	22.30$\pm$2.35	&	2.54$\pm$0.46	&	10-11	\\
820123	&	0.2198		&	15.79$\pm$5.18	&	7.45$\pm$3.87	&	10-11	\\
811924	&	0.2204		&	24.11$\pm$5.74	&	14.57$\pm$5.28	&	10-11	\\
824857	&	0.1859		&	45.74$\pm$4.85	&	15.56$\pm$2.69	&	10-11	\\
833897	&	0.2639		&	12.45$\pm$2.17	&	3.24$\pm$0.96	&	10-11	\\
846186	&	0.2500		&	20.58$\pm$5.66	&	13.57$\pm$4.64	&	10-11	\\
825188	&	0.2510		&	19.34$\pm$4.26	&	5.75$\pm$1.44	&	10-11	\\
845491	&	0.2501		&	12.18$\pm$2.57	&	6.20$\pm$1.84	&	10-11	\\
\enddata 
\end{longtable}

\clearpage

%

%

%

\clearpage

%

%

%

%

%

%

%

%

%

%

%

%

%

%

\clearpage

%

%

%

%

%

%

%

\end{document}